\documentclass{article} 
\usepackage{iclr2027_conference,mathpazo,times}
\usepackage{microtype}

\usepackage{amsmath,amsfonts,bm}

\def\eqref#1{equation~\ref{#1}}

\def\1{\bm{1}}

\DeclareMathAlphabet{\mathsfit}{\encodingdefault}{\sfdefault}{m}{sl}
\SetMathAlphabet{\mathsfit}{bold}{\encodingdefault}{\sfdefault}{bx}{n}

\usepackage{float}
\usepackage[hidelinks]{hyperref}
\hypersetup{
  pdftitle={TaoD2C-Bench: Benchmarking MLLMs for Industrial UI Code Generation Beyond Visual Fidelity},
  pdfauthor={Chengwei Shi, Yunnong Chen, Tingting Zhou, Qiang Lu, Shiyu Yue, Xinyuan Hu, Jianfang Ru, Liuqing Chen}
}
\usepackage{url}
\usepackage{booktabs}
\usepackage{multirow}
\usepackage{graphicx}
\usepackage{pifont}
\usepackage{amssymb}
\usepackage{booktabs}
\usepackage{multirow}
\usepackage[table]{xcolor}
\definecolor{rankhi}{HTML}{6C5CE7}
\definecolor{ranklo}{HTML}{F5B301}
\definecolor{taskblue}{HTML}{2D5F9A}
\definecolor{taskorange}{HTML}{A84D13}
\definecolor{taskgreen}{HTML}{3E7136}
\newcommand{\taskcapability}[1]{%
  \mbox{}\par\nobreak\smallskip
  \begingroup
  \setlength{\fboxsep}{4pt}%
  \noindent\colorbox{black!6}{%
    \parbox{\dimexpr\linewidth-2\fboxsep\relax}{\textit{Capability:} #1}%
  }\par
  \endgroup
  \nobreak\smallskip
  \noindent\ignorespaces
}
\newcommand{\rhi}[1]{\cellcolor{rankhi!#1}}
\newcommand{\rlo}[1]{\cellcolor{ranklo!#1}}
\usepackage{tikz}
\newcommand{\cnum}[1]{\raisebox{-0.3pt}{\tikz[baseline=(c.base)]{\node[circle,fill=black!45,text=white,inner sep=0pt,minimum size=0.95em,font=\tiny\bfseries](c){#1};}}}

\usepackage[a4paper,margin=2.5cm,headheight=32pt,headsep=25pt]{geometry}
\definecolor{headerink}{HTML}{4D4D4D}
\definecolor{headerrule}{HTML}{4D4D4D}
\definecolor{referenceblue}{rgb}{0.15,0.45,0.68}
\definecolor{linkblue}{rgb}{0,0,0.5}
\hypersetup{colorlinks=true,citecolor=referenceblue,linkcolor=linkblue,urlcolor=linkblue}

\definecolor{brandorange}{HTML}{FF6A00}
\newcommand{\taodlogo}[2]{%
  \raisebox{-0.5\height}{\includegraphics[height=#1]{figures/#2}}%
}
\renewcommand{\headrule}{%
  {\color{headerrule}\hrule height 0.8pt width \headwidth}%
  \vskip -0.8pt
}
\fancypagestyle{taodtitle}{%
  \fancyhead{}%
  \fancyhead[L]{%
    \taodlogo{14pt}{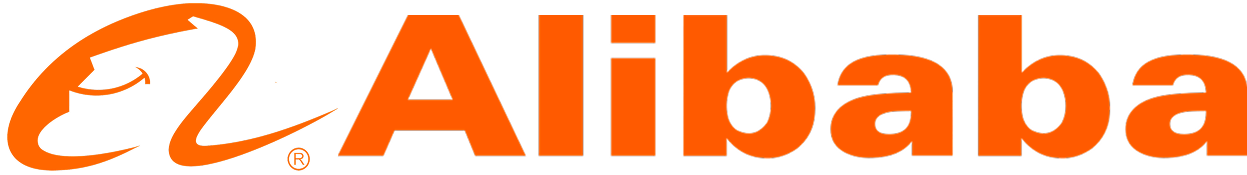}\hspace{22pt}%
    \taodlogo{20pt}{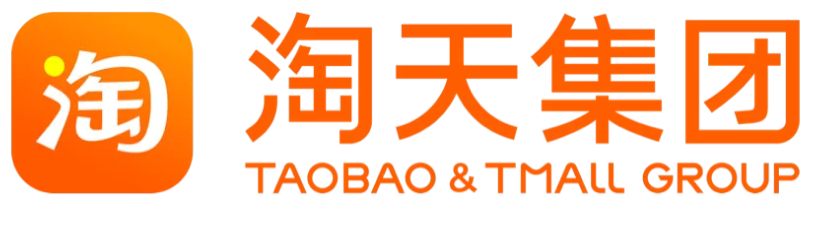}}%
  \fancyhead[R]{\taodlogo{23pt}{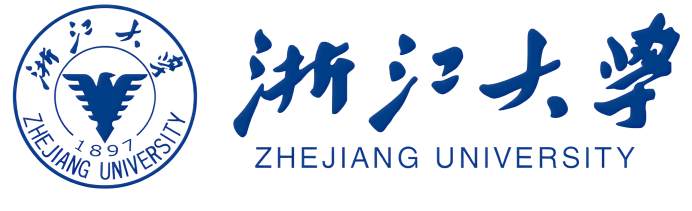}}%
}
\fancypagestyle{taodrunning}{%
  \fancyhead{}%
  \fancyhead[L]{%
    \taodlogo{7.5pt}{ali_logo.png}\hspace{12pt}%
    \taodlogo{10.5pt}{tao_logo.png}}%
  \fancyhead[C]{{\sffamily\fontsize{8}{10}\selectfont\color{headerink}%
    \textbf{TaoD2C-Bench}\enspace\textbar\enspace Under review}}%
  \fancyhead[R]{\taodlogo{12pt}{zju_logo.png}}%
}

\title{\textcolor{brandorange}{\texttt{TaoD2C-Bench}}: Benchmarking MLLMs for\\
Industrial UI Code Generation Beyond Visual Fidelity}

\usepackage{latexml}
\iflatexml
\usepackage{authblk}
\let\taodnativefootnote\footnote
\newcommand{\taodauthornote}[2]{\taodnativefootnote[#1]{#2}}
\author[1]{Chengwei Shi\taodauthornote{\textdagger}{Work done during internship at Alibaba.}}
\author[2,1]{Yunnong Chen\taodauthornote{*}{Corresponding author: \href{mailto:chen_yn@zju.edu.cn}{\texttt{chen\_yn@zju.edu.cn}}.}}
\author[2]{Tingting Zhou}
\author[2]{Qiang Lu}
\author[2]{Shiyu Yue}
\author[2]{Xinyuan Hu}
\author[2]{Jianfang Ru}
\author[1]{Liuqing Chen\taodauthornote{*}{Corresponding author: \href{mailto:chenlq@zju.edu.cn}{\texttt{chenlq@zju.edu.cn}}.}}
\affil[1]{Zhejiang University}
\affil[2]{Taobao \& Tmall Group of Alibaba}
\else
\author{
Chengwei Shi$^{1,\dagger}$ \quad Yunnong Chen$^{2,1,*}$ \quad Tingting Zhou$^{2}$ \quad Qiang Lu$^{2}$ \\[2pt]
Shiyu Yue$^{2}$ \quad Xinyuan Hu$^{2}$ \quad Jianfang Ru$^{2}$ \quad Liuqing Chen$^{1,*}$ \\[3pt]
$^{1}$\textbf{Zhejiang University} \qquad $^{2}$\textbf{Taobao \& Tmall Group of Alibaba} \\[2pt]
{\small\texttt{chen\_yn@zju.edu.cn} \qquad \texttt{chenlq@zju.edu.cn}}
}
\fi

\iclrfinalcopy 
\begin{document}

\maketitle
\pagestyle{taodrunning}
\thispagestyle{taodtitle}

\iflatexml\else
\begingroup
\makeatletter
\long\def\@makefntext#1{\noindent#1}
\makeatother
\renewcommand{\thefootnote}{}
\begin{NoHyper}
\footnotetext{%
  $^{\dagger}$Work done during internship at Alibaba.\\
  $^{*}$Corresponding authors.
}
\end{NoHyper}
\endgroup
\fi

\begin{abstract}
A key challenge for multimodal large language models (MLLMs) is moving beyond visual recognition to constraint-aware cross-modal reasoning. This involves combining visual cues with information from other modalities to understand elements' relationships under domain-specific rules. This challenge is acutely evident in industrial design-to-code (D2C), which converts user interface (UI) designs into code and requires MLLMs to connect design images with disorganized layer metadata, infer component and layout implementation requirements, and realize them in code under target-library constraints. However, these capabilities remain insufficiently evaluated in realistic industrial settings. To fill this gap, we present TaoD2C-Bench, a benchmark for evaluating MLLMs' ability to generate UI code that satisfies implementation requirements in industrial applications. The TaoD2C dataset consists of 2,861 production designs from 17 commercial platforms with 97,652 expert annotations across four categories: Component, Group, Alignment, and Position. These annotations distinguish required constraints from permitted implementation choices. TaoD2C-Bench defines three tasks: end-to-end UI code generation, requirement inference, and requirement realization. Evaluating eight MLLMs reveals substantial gaps in generating UI code that satisfies implementation requirements, alongside distinct performance profiles in inference and realization. We further show that MLLMs' visual reconstruction ability does not necessarily imply an ability to generate code that meets these requirements. We release TaoD2C to support research on industrial UI code generation (\url{https://taod2c-bench.github.io/}).
\end{abstract}

\begin{figure}[!ht]
\centering
\includegraphics[width=\linewidth]{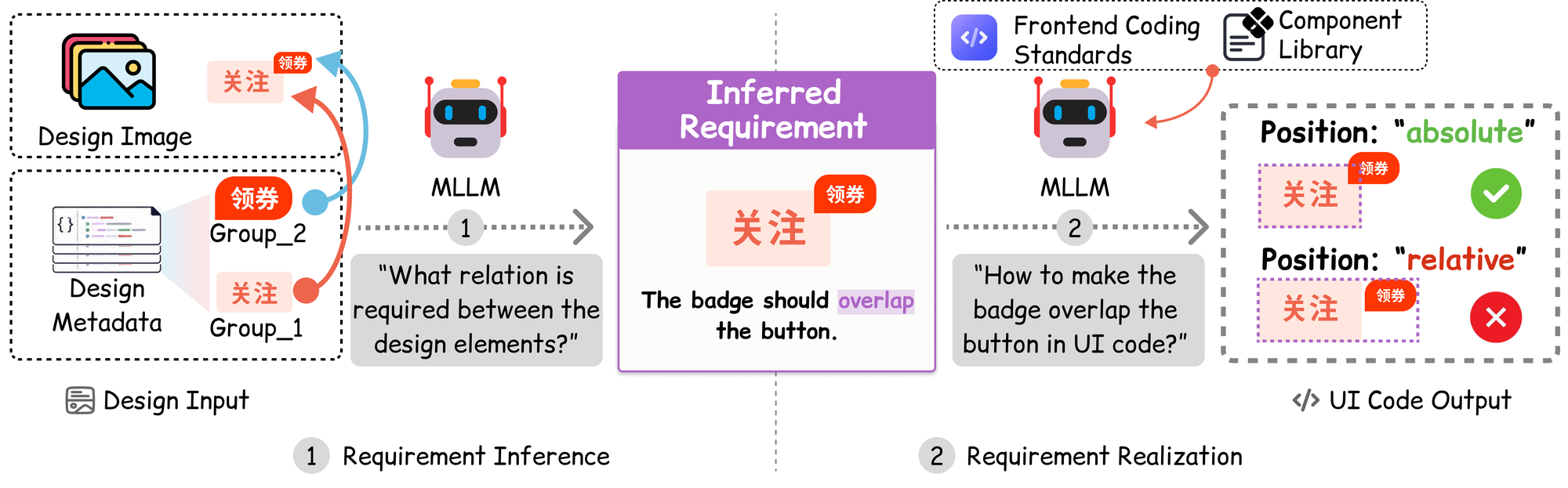}
\caption{Two capabilities in UI code generation: (1) inferring positioning
requirements and (2) realizing them in code.}
\label{fig:stages}
\end{figure}

\section{Introduction}

Industrial design-to-code (D2C) aims to automate frontend development by translating UI designs
into executable code \citep{xu2025webvia,xiao2025interaction2code}.
Although multimodal large language models (MLLMs) have improved the visual fidelity of generated
interfaces \citep{si2025design2code}, visually similar outputs can still fail to meet the
implementation requirements of the specific design, such as which components and properties to use
and how elements should be grouped, aligned, and positioned, or to follow frontend coding standards
\citep{reactRules,googleHtmlCssGuide,airbnbReactGuide}. Such generated code can degrade user experience
and require manual rework before deployment
\citep{moran2018gvt,walsh2017layout,mahajan2015websee}.

As illustrated in Figure~\ref{fig:stages}, generating code that meets these implementation
requirements depends on two MLLM capabilities: inferring them from designs and realizing them
in code. Inference requires cross-modal reasoning to associate disorganized layer metadata
with UI images despite visual complexity and limited semantic cues \citep{chen2024egfe}.
Realization requires generating code under complex constraints imposed by these requirements
and frontend coding standards.
Generated code alone cannot reveal which of these two capabilities limits MLLM performance in UI code generation. Evaluating them requires datasets that pair UI designs with their implementation requirements.

However, datasets pairing industrial UI designs with implementation requirements remain
scarce. Existing evaluations also have three limitations.
\textbf{(a) Visual evaluation} compares rendered appearance but does not verify whether the
code uses the intended components and layout structure \citep{xu2026onedbench}.
\textbf{(b) Heuristic evaluation} favors coding practices that may conflict with a design's
implementation requirements. For example, Figma2Code favors a lower absolute-positioning
ratio, even when absolute positioning is appropriate for overlapping elements
\citep{gui2026figma2code}.
\textbf{(c) Single-reference structural comparison} may penalize valid code that differs from
the reference, overlooking multiple implementations that satisfy the same requirements
\citep{guo2025iwbench}.

\begin{figure}[t]
\centering
\includegraphics[width=\linewidth]{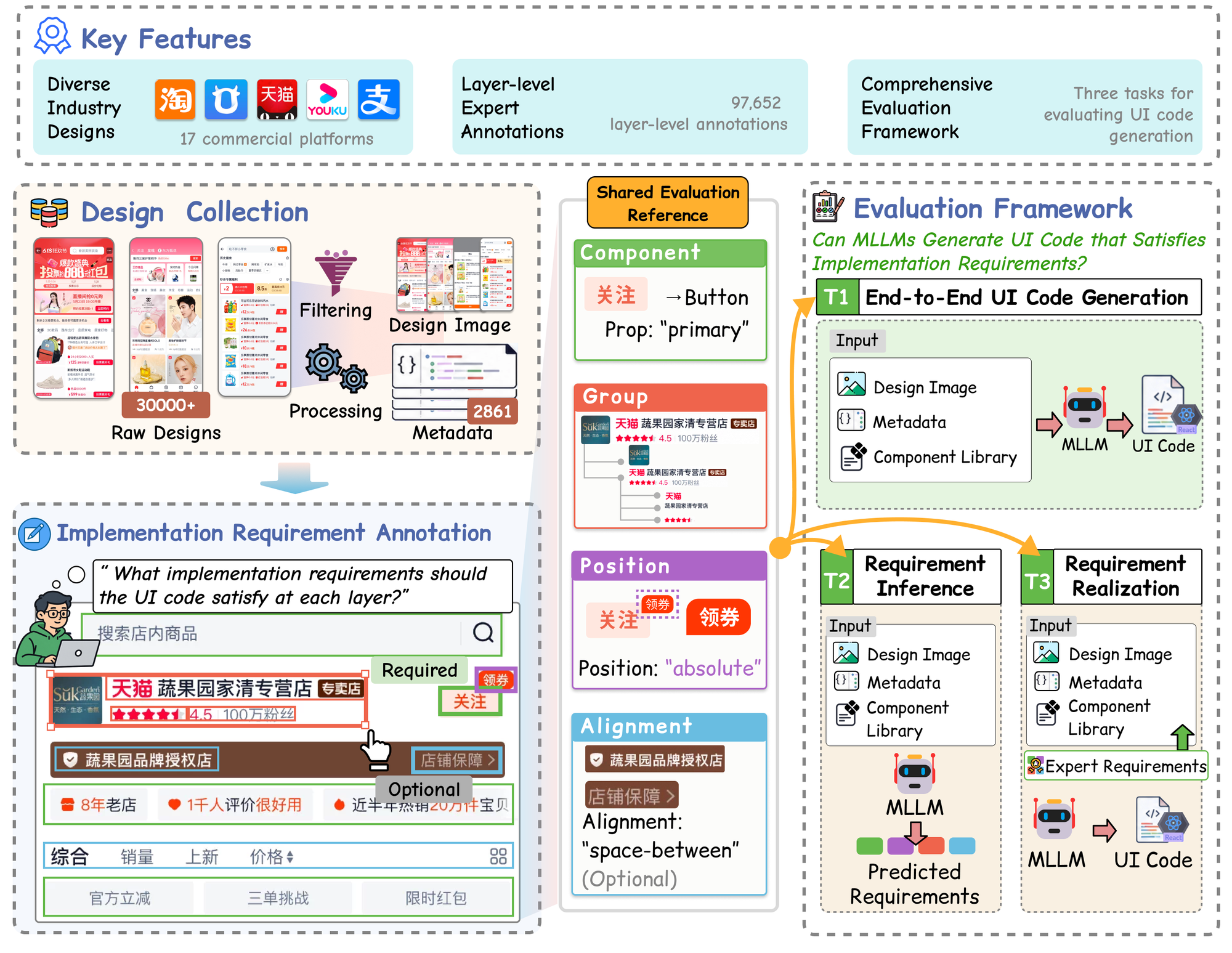}
\caption{Overview of TaoD2C-Bench. We collect 2,861 production UI designs in TaoD2C and annotate their implementation requirements across Component, Group, Alignment, and Position. These annotations provide a shared evaluation reference for UI code generation (T1), requirement inference (T2), and realization with expert requirements supplied (T3).}
\label{fig:overview}
\end{figure}

To fill this gap, we introduce TaoD2C-Bench (Figure~\ref{fig:overview}), a benchmark for evaluating
MLLMs' ability to generate UI code that satisfies implementation requirements in industrial
applications. We collect 2,861 production UI designs from 17 commercial platforms, spanning
diverse consumer-facing and business-facing scenarios. Each design includes an image, layer
metadata, and its target component library. To associate designs with implementation requirements, we
collaborated with frontend engineering experts to produce 97,652 layer-level annotations across
four key categories: \emph{Component}, \emph{Group}, \emph{Alignment}, and \emph{Position}.
Required annotations specify requirements for the generated code, while optional annotations
describe implementation choices that are allowed but not required. These annotations provide a
shared evaluation reference for three tasks. \emph{End-to-End UI Code Generation} (T1) evaluates
MLLMs' overall ability to generate code that satisfies implementation requirements.
\emph{Requirement Inference} (T2) and \emph{Requirement Realization} (T3) separately assess
MLLMs' abilities to infer implementation requirements and realize them in code, helping
distinguish limitations in these two capabilities.

Our experiments with eight MLLMs reveal notable limitations in their ability to generate UI
code that satisfies implementation requirements, alongside distinct performance profiles in
requirement inference and realization. We further show that MLLMs' visual reconstruction
ability does not necessarily imply an ability to generate code that meets these requirements.
Our analysis highlights the need for design-specific evaluation: implementation requirements
vary across designs, and a single design can admit multiple valid implementations. These
insights provide a foundation for future research on improving MLLMs' ability to generate UI
code.

Our main contributions are as follows:
\begin{itemize}
\item \textbf{Expert-Annotated Industrial Dataset.} We introduce TaoD2C, a large-scale,
high-quality industrial UI dataset comprising 2,861 designs. It pairs production designs from
17 commercial platforms across consumer-facing and business-facing applications with expert
annotations of layer-level implementation requirements, ensuring industrial realism, application
diversity, and access to expert knowledge.
\item \textbf{Implementation-Requirement Evaluation Framework.} We develop TaoD2C-Bench, an
 evaluation framework based on expert-annotated implementation requirements across four key
categories. Its three tasks evaluate MLLMs' abilities to generate UI code that satisfies
these requirements, infer them from designs, and realize them in code when supplied.
\item \textbf{Empirical Findings.} Through extensive experiments on eight MLLMs, we reveal
critical failures in generating UI code that satisfies implementation requirements,
while characterizing MLLM performance profiles in requirement inference and realization.
We further show that MLLMs' visual reconstruction ability does not necessarily imply
an ability to generate code that meets these requirements.
\end{itemize}

\section{Related Work}

\paragraph{UI-to-Code Generation.}
Early UI-to-code systems reconstructed interfaces from screenshots
\citep{nguyen2015remaui,beltramelli2018pix2code,chen2018ui2code,moran2018redraw}.
Recent MLLM-based methods employ synthetic data and feedback-based finetuning \citep{laurencon2024websight,ge2025flame,wu2024uicoder,zheng2026unicoder}.
Decomposition and layout reasoning organize complex interfaces into manageable regions
\citep{wan2025dcgen,wu2025layoutcoder,gui2025latcoder,jiang2025screencoder,zhou2025declarui}.
Hierarchical generation \citep{gui2025uicopilot,chen2025designcoder} and iterative visual refinement
\citep{yue2025uiorchestra,sansford2026visionguided,yang2025ui2coden,deng2026visrefiner,xiong2026rubse} further improve reconstruction.
Beyond image-only generation, recent work leverages multimodal design inputs,
including images, metadata, and assets, to support higher-quality UI code generation
\citep{xiao2024prototype2code}.
Overall, despite these advances, generating production-ready UI code for industrial
applications remains challenging \citep{xiao2026efficientuicoder}.

\paragraph{UI Code Generation Evaluation.}
UI code generation evaluation considers appearance, coding practices, and structure.
\emph{Visual evaluation} compares rendered outputs with target images through image and
layout similarity \citep{xiao2025designbench,lai2025webrenderbench,li2025sketch2code,zhang2026widget2code},
using metrics such as SSIM, CLIP, and LPIPS \citep{wang2004ssim,radford2021clip,zhang2018lpips}.
WebUIBench extends evaluation to element attributes and spatial layouts \citep{lin2025webuibench}.
\emph{Heuristic evaluation}, exemplified by Figma2Code, summarizes practices such as
relative-unit and absolute-positioning usage as proxies for code quality \citep{gui2026figma2code}.
\emph{Single-reference structural comparison} assesses agreement with a reference
implementation \citep{sun2025fullfront,xiao2026comuicoder}.
However, visual similarity does not establish requirement satisfaction.
Aggregate coding statistics cannot determine whether a practice suits a specific design,
while differences from one reference can reflect valid implementations.
Overall, evaluation has expanded beyond appearance to code quality, but assessing
implementation requirement satisfaction while accommodating multiple valid implementations
remains an open challenge.

\raggedbottom
\section{TaoD2C-Bench}
\label{sec:benchmark}

\subsection{Dataset Construction}

\begin{figure}[t]
\centering
\begin{minipage}[t]{0.49\linewidth}
\centering
\includegraphics[width=\linewidth,height=0.863636\linewidth,keepaspectratio]{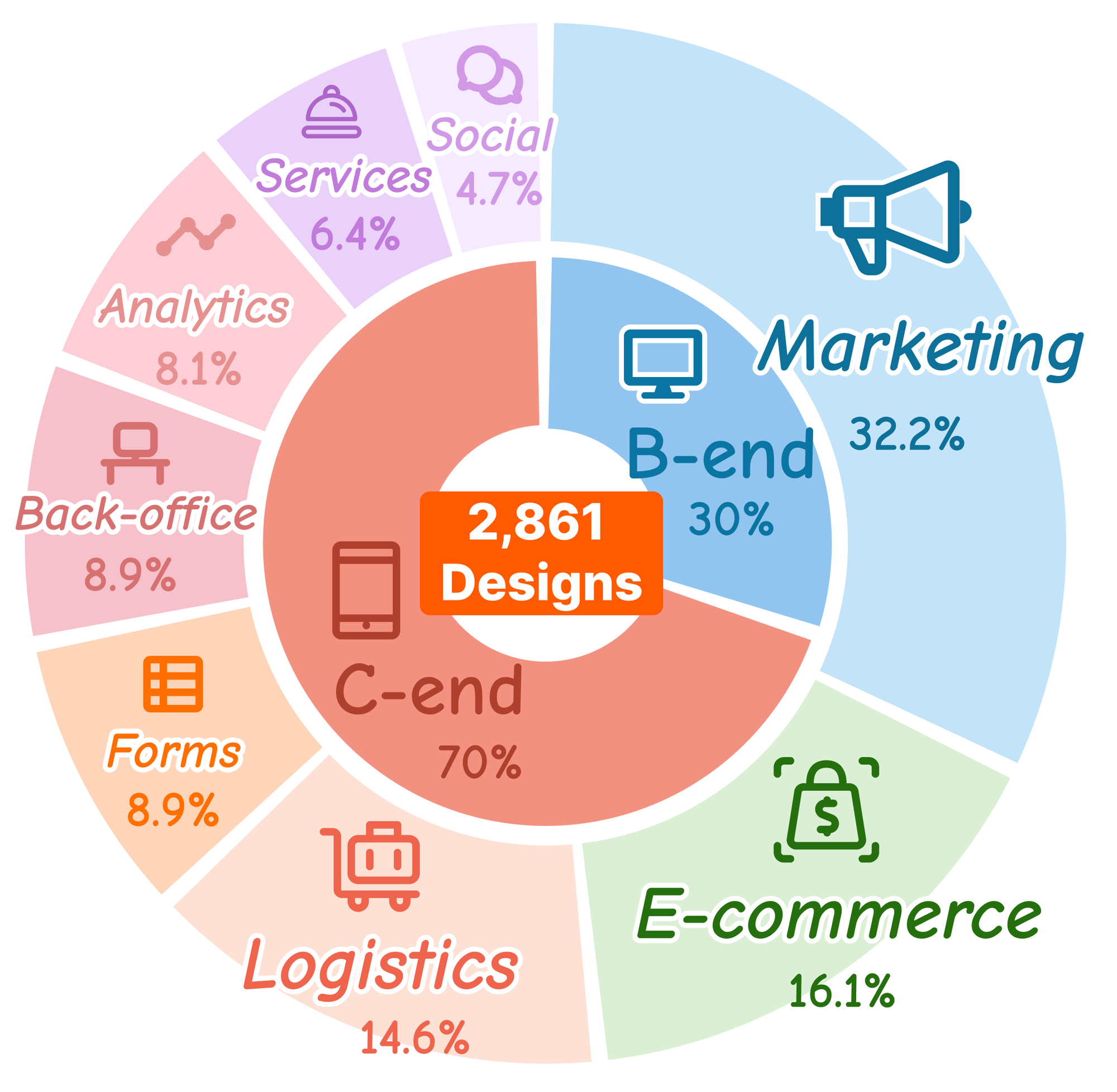}
\caption{TaoD2C dataset composition by endpoint type and application scenario.}
\label{fig:dataset-overview}
\end{minipage}
\hfill
\begin{minipage}[t]{0.49\linewidth}
\centering
\includegraphics[width=\linewidth,trim=0 18bp 0 0,clip]{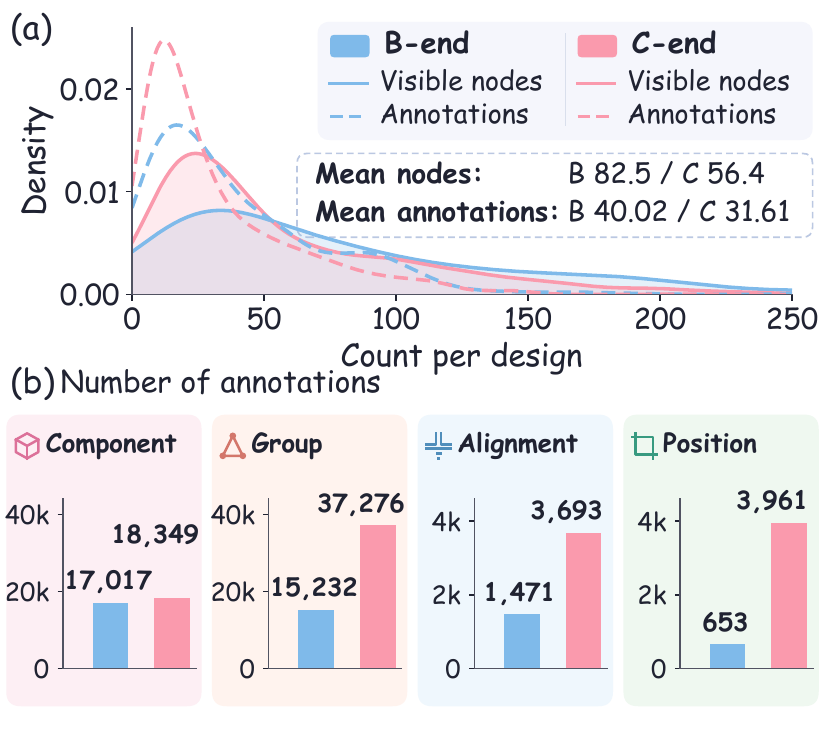}
\caption{Per-design node and annotation distributions (a) and annotation counts by category (b).}
\label{fig:dataset-distributions}
\end{minipage}
\end{figure}

We constructed TaoD2C through four stages: design collection, filtering and normalization,
expert annotation, and evaluation subset selection. Figure~\ref{fig:dataset-overview} summarizes the
dataset composition, Figure~\ref{fig:dataset-distributions} reports the per-design complexity
and annotation counts, and Appendix~\ref{app:platform} provides the full construction procedure.

\paragraph{Design collection.}
To capture the complexity of real-world industrial UIs, we collected MasterGo design files
from multiple product teams within a large-scale e-commerce ecosystem. With authorization
from the owning teams, we assembled an initial pool of approximately 30,000 design pages
from 17 commercial platforms. These designs span consumer-facing (C-end) interfaces for
shopping, marketing, membership, content, and transaction flows, as well as business-facing
(B-end) interfaces for merchant operations, data analytics, logistics management, and complex
business workflows. For each design, we extracted its rendered image and layer metadata and
retained a reference to its target component library. We then filtered, normalized, and
quality-checked the extracted data to ensure data quality and remove duplicates. This collection
captures diverse interface structures and business workflows encountered in industrial production.

\paragraph{Implementation Requirement Annotation.}
To annotate the collected designs, we developed a platform that integrates their rendered images
and parsed layer trees, allowing experts to select individual layers and annotate their requirements. Two senior frontend engineers with seven and five years of experience, respectively,
annotated the designs. They inspected the source design, rendered image, layer metadata, and
target component library documentation to specify component types and properties, container
hierarchies, alignment relations, and positioning schemes. The annotations were based on W3C CSS
layout and positioning specifications \citep{w3cFlexbox,w3cPosition}, target-library definitions and APIs, and the annotators'
understanding of each design and frontend implementation experience. Each annotation is marked as
\textbf{required} or \textbf{optional}. Required annotations specify constraints that an
implementation must satisfy, while optional annotations record permitted implementation choices. A third expert iteratively reviewed
all annotations, with identified errors corrected by the frontend experts and rechecked.
The annotation and verification process spanned
18 months. We denote the four requirement categories as \textbf{Component},
\textbf{Group}, \textbf{Alignment}, and \textbf{Position} throughout the tables and analysis:

\begin{enumerate}
\item \textbf{Component}: component implementation, defined by the target-library component
type and property configuration. For a set of layers that realizes a reusable component,
annotators identify its type in the target library and configure the properties defined by
its API, such as a switch state, button size,
input placeholder, or a select's options and selected value.
\item \textbf{Group}: container nesting relations. Annotators mark the container regions,
and member nodes are deterministically compiled from those regions and the layer tree.
\item \textbf{Alignment}: layout relations. For containers with specific alignment
requirements, annotators identify the rule arranging their children along the main axis, such
as \texttt{space-between}, \texttt{center}, \texttt{flex-start}, \texttt{flex-end}, and
other \texttt{justify-content} values.
\item \textbf{Position}: positioning schemes. Annotators specify positioning schemes for
individual elements, such as \texttt{absolute} positioning relative to a containing block.
\end{enumerate}

\subsection{Dataset Characteristics}

\begin{table}[tbp]
\centering
\footnotesize
\setlength{\tabcolsep}{4pt}
\caption{Comparison with existing design-to-code benchmarks.
Implementation requirements are annotated by experts.}
\label{tab:datasets}
\begin{tabular*}{\linewidth}{@{\extracolsep{\fill}}>{\raggedright\arraybackslash}p{4.15cm}llccl@{}}
\toprule
Benchmark & Data source & Scale & Metadata & \shortstack{Implementation\\requirements} & Granularity \\
\midrule
Design2Code \citep{si2025design2code} & Web pages & 484 & \ding{55} & \ding{55} & Page \\
Web2Code \citep{yun2024web2code} & Synthetic+web & 884.7K & \ding{55} & \ding{55} & Page/QA \\
WebUIBench \citep{lin2025webuibench} & Websites & 21.8K QA & \ding{55} & \ding{55} & QA/element \\
IW-Bench \citep{guo2025iwbench} & Generated+web & 1,200 & \ding{55} & \ding{55} & DOM/layout \\
WebCode2M \citep{gui2025webcode2m} & Common Crawl & 2.56M & \ding{55} & \ding{55} & Element/layout \\
Figma2Code \citep{gui2026figma2code} & Figma Comm. & 3,055 (213 eval) & \checkmark & \ding{55} & Page \\
1D-Bench \citep{xu2026onedbench} & Industrial & 984 (204 eval) & \checkmark & \ding{55} & Instance \\
\midrule
\textbf{TaoD2C-Bench (ours)} & Industrial & 2,861 & \checkmark & \checkmark & Layer/relation \\
\bottomrule
\end{tabular*}
\end{table}

\paragraph{\protect\cnum{1}~Industrial realism and complexity.}
TaoD2C captures the complexity of industrial UI development through 2,861 production designs
from 17 commercial platforms spanning eight application scenarios (Figure~\ref{fig:dataset-overview}).
These designs encompass complex business forms and tables alongside image-rich consumer interfaces,
averaging 64.2 visible layers per design. This structural
and application diversity provides a challenging testbed for evaluating MLLMs under the
heterogeneous component and layout requirements encountered in production.

\paragraph{\protect\cnum{2}~Dense expert annotations.}
TaoD2C combines design images, layer metadata, and target component libraries with expert
annotations of implementation requirements for layers and their relations, a combination unique among the compared
benchmarks (Table~\ref{tab:datasets}). Its 97,652 annotations span Component, Group, Alignment,
and Position, distinguishing required choices from optional alternatives. Building this resource
involved 18 months of expert annotation and iterative verification. Drawing on human frontend
expertise, these annotations specify design-specific component and layout requirements for
code implementation. They provide a shared reference for evaluating MLLMs' abilities to infer
these requirements and generate UI code that satisfies them.

\section{Tasks and Metrics}
\label{sec:tasks}

\subsection{Tasks}

\begin{figure}[t]
\centering
\includegraphics[width=\linewidth]{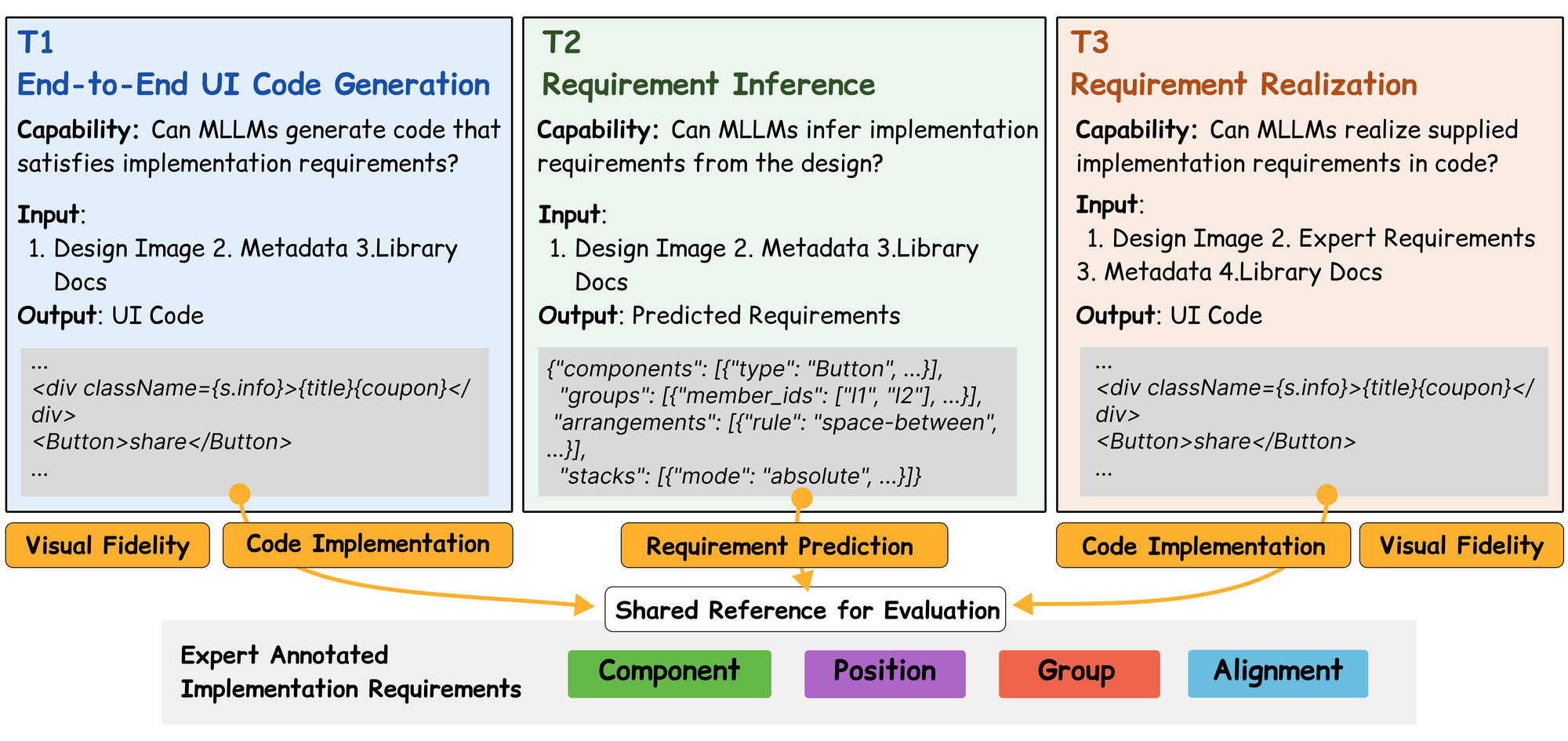}
\caption{Three tasks and their evaluation. Expert annotations provide a shared reference for
requirement prediction (T2) and code implementation (T1/T3); visual fidelity compares rendered
pages with design images.}
\label{fig:task-metrics}
\end{figure}

We define three tasks to evaluate end-to-end UI code generation, requirement inference,
and requirement realization (Figure~\ref{fig:task-metrics}). All three tasks use the same designs
and expert annotations as a shared evaluation reference.

\paragraph{T1: End-to-End UI Code Generation.}
\taskcapability{\textcolor{taskblue}{\textbf{Generating UI code that satisfies implementation requirements}}
while reproducing the design's appearance.}
Given a design image, layer metadata, and target-library documentation, the MLLM generates
executable UI code. We assess requirement satisfaction through source-code and runtime analysis,
and evaluate visual fidelity separately.

\paragraph{T2: Requirement Inference.}
\taskcapability{\textcolor{taskgreen}{\textbf{Inferring implementation requirements}} from a design image,
layer metadata, and target-library documentation.}
Given the same inputs as T1, the MLLM predicts structured requirements across Component, Group,
Alignment, and Position without generating code. We compare these predictions with expert
annotations to evaluate requirement inference separately from code implementation.

\paragraph{T3: Requirement Realization.}
\taskcapability{\textcolor{taskorange}{\textbf{Realizing supplied implementation requirements}} in UI code
under complex constraints imposed by these requirements and frontend coding standards.}
To evaluate MLLMs' requirement realization ability, we supply the T1 inputs together with
the expert annotations for the input design.
The MLLM generates executable UI code from these inputs. We assess whether the generated
code realizes the supplied requirements and evaluate visual fidelity using the same metrics
as T1.

\subsection{Metrics}

We evaluate the predicted requirements from T2 and the generated code from T1/T3 against
shared expert annotations. We assess visual fidelity by comparing pages rendered from T1/T3
code with the design images (Figure~\ref{fig:task-metrics}).
We account for required and optional annotations when computing the scores, with details
provided in Appendix~\ref{app:metrics}.

\paragraph{Requirement Prediction.}
For T2, we compare predicted requirements $\hat R$ with expert annotations $R^*$ through
layer-level correspondences, checking both the associated design elements and category-specific
attributes. This evaluates requirement inference separately from code implementation.
We report micro-F1 for Component, Group, Alignment, and Position.
\begin{equation}
\label{eq:score-pred}
\mathrm{Score}_{\mathrm{pred}} = \mathcal{E}_{\mathrm{pred}}(\hat R, R^*).
\end{equation}

\paragraph{Code Implementation.}
For T1 and T3, we bridge design-side annotations $R^*$ and runtime implementations by
establishing requirement-specific correspondences using source-code, structural, and geometric
evidence. This enables requirement-level evaluation of generated code $C$ without reference code.
We report micro-F1 for Type, Props, Group, Alignment, and Position; Props is evaluated only on
type-matched components.
\begin{equation}
\label{eq:score-impl}
\mathrm{Score}_{\mathrm{impl}} = \mathcal{E}_{\mathrm{impl}}\!\left(\Phi(C), R^*\right).
\end{equation}
Here, $\Phi(C)$ extracts source-code, structural, and geometric evidence, which
$\mathcal{E}_{\mathrm{impl}}$ matches against $R^*$ to compute per-dimension scores.

\paragraph{Visual Fidelity.}
For T1 and T3, we compare $\mathrm{Render}(C)$ with the design image $I$
\citep{si2025design2code,xu2026onedbench}. We report 1D-Bench's composite visual score (CV),
visual element similarity (VES) using DINOv2 features \citep{oquab2023dinov2}, and normalized
RGB mean absolute error (MAE). Higher CV/VES and lower MAE indicate better visual fidelity.
Appendix~\ref{app:visual} provides computation details.
\begin{equation}
\label{eq:score-vis}
\mathrm{Score}_{\mathrm{vis}} = \mathcal{E}_{\mathrm{vis}}(\mathrm{Render}(C), I).
\end{equation}

\section{Experiments}
\label{sec:experiments}

\providecommand{\dd}[1]{{\scriptsize\,(#1)}}
\providecommand{\stk}[2]{\shortstack{#1\\[1pt]{\tiny #2}}}

\subsection{Setup}

We evaluate eight MLLMs on a curated test set of 421 TaoD2C designs, comprising 152 B-end
and 269 C-end designs, with all selected designs undergoing annotation review and deduplication.
All MLLMs receive the same
design context and task prompts. The evaluated MLLMs are GPT-5.6 Terra, GPT-5.4, Claude Sonnet 5,
Qwen3.8-Max, Qwen3-VL-235B-A22B-Thinking, Gemini 3.5 Flash, Kimi K3, and Gemini 3.1 Pro
Preview. Further experimental details are provided in the appendix.

\subsection{Results}

Table~\ref{tab:main} reports Code Implementation and Visual Fidelity scores for T1 and
Requirement Prediction scores for T2 on TaoD2C-Bench, including per-MLLM results and averages
across MLLMs. Table~\ref{tab:guided} reports Code Implementation and Visual Fidelity scores
for Requirement Realization (T3). Values below each score show the change from T1.

\begin{table*}[t]
\setlength{\belowcaptionskip}{9pt}
\caption{Main results for T1 and T2 on 421 designs. Code Implementation and Requirement Prediction scores are micro F1 (\%).
Higher CV/VES and lower MAE indicate better visual fidelity. Bold denotes the best
result; shading indicates within-column rank.}
\label{tab:main}
\centering

\small
\setlength{\tabcolsep}{2.0pt}
\renewcommand{\arraystretch}{1.15}

\begin{tabular*}{\textwidth}{@{\extracolsep{\fill}}lcccccccccccc@{}}
\toprule
\multirow{3}{*}{Model}
& \multicolumn{8}{c}{\textbf{End-to-End UI Code Generation (T1)}}
& \multicolumn{4}{c}{\multirow{2}{*}{\shortstack{\textbf{Requirement}\\\textbf{Inference (T2)}}}}
\\
\cmidrule(lr){2-9}
& \multicolumn{5}{c}{\textbf{Code Implementation}}
& \multicolumn{3}{c}{\textbf{Visual Fidelity}}
& \multicolumn{4}{c}{}
\\
\cmidrule(lr){2-6}
\cmidrule(lr){7-9}
\cmidrule(lr){10-13}
& Type & Props & Group & Align. & Pos.
& CV $\uparrow$ & VES $\uparrow$ & MAE $\downarrow$
& Comp. & Group & Align. & Pos.
\\
\midrule

GPT-5.6 Terra
& \rhi{20} 62.47
& \rhi{34} \textbf{44.25}
& \rhi{9} 56.84
& \rhi{34} \textbf{30.53}
& \rhi{9} 36.43
& \rhi{34} \textbf{85.78}
& \rhi{9} 0.9174
& \rhi{9} 0.1163
& \rhi{20} 59.15
& 60.92
& \rhi{34} \textbf{37.11}
& 47.68
\\

GPT-5.4
& 60.19
& 40.29
& 56.48
& \rhi{20} 27.48
& 33.44
& \rhi{9} 85.09
& \rhi{20} 0.9183
& \rhi{34} \textbf{0.1078}
& \rlo{22} 57.32
& 60.93
& 31.88
& \rhi{9} 48.85
\\

Claude Sonnet 5
& \rlo{12} 59.89
& \rhi{9} 40.37
& \rlo{12} 54.82
& \rlo{12} 23.54
& \rlo{22} 27.93
& \rlo{22} 78.82
& \rlo{22} 0.8946
& \rlo{22} 0.1350
& \rlo{12} 58.36
& \rlo{22} 50.48
& \rhi{20} 34.84
& \rlo{12} 47.62
\\

Qwen3.8-Max
& 61.77
& 38.83
& \rhi{34} \textbf{59.46}
& 26.42
& \rhi{34} \textbf{45.35}
& \rhi{20} 85.53
& \rhi{34} \textbf{0.9192}
& \rhi{20} 0.1134
& \rhi{9} 58.86
& \rhi{34} \textbf{71.40}
& \rlo{12} 29.45
& \rhi{20} 49.15
\\

Qwen3-VL-235B
& \rlo{36} 31.60
& \rlo{36} 17.34
& \rlo{36} 30.53
& \rlo{36} 6.72
& \rlo{36} 7.61
& \rlo{36} 58.37
& \rlo{36} 0.7919
& \rlo{36} 0.1808
& \rlo{36} 53.25
& \rlo{36} 21.31
& \rlo{36} 17.68
& \rlo{36} 26.75
\\

Gemini 3.5 Flash
& \rhi{9} 62.07
& \rlo{12} 38.03
& \rlo{22} 51.28
& \rlo{22} 20.10
& \rlo{12} 29.49
& \rlo{12} 80.50
& \rlo{12} 0.9112
& \rlo{12} 0.1195
& 58.80
& \rlo{12} 51.42
& \rhi{9} 34.61
& \rlo{22} 47.55
\\

Kimi K3
& \rhi{34} \textbf{63.25}
& \rhi{20} 41.56
& \rhi{20} 59.41
& \rhi{9} 27.04
& \rhi{20} 43.68
& 84.40
& 0.9151
& 0.1178
& \rhi{34} \textbf{59.67}
& \rhi{20} 70.84
& 33.31
& \rhi{34} \textbf{51.93}
\\

Gemini 3.1 Pro
& \rlo{22} 57.74
& \rlo{22} 36.87
& 55.52
& 24.59
& 33.81
& 83.00
& 0.9155
& 0.1166
& 58.67
& \rhi{9} 67.96
& \rlo{22} 20.64
& 48.28
\\

\midrule
Mean
& 57.37
& 37.19
& 53.04
& 23.30
& 32.22
& 80.19
& 0.8979
& 0.1259
& 58.01
& 56.91
& 29.94
& 45.98
\\

\bottomrule
\end{tabular*}
\end{table*}

\begin{samepage}
\paragraph{Finding 1.} \emph{Current MLLMs exhibit limited capabilities in generating UI
code that satisfies implementation requirements.}
Even the best end-to-end results in Table~\ref{tab:main} reach only 63.25 F1 on component
type (Kimi K3), 44.25 on properties (GPT-5.6 Terra), and 59.46 on grouping (Qwen3.8-Max).
In the overall T1 results, Alignment scores lowest for all eight MLLMs, averaging 23.30 F1
and a best score of 30.53 (GPT-5.6 Terra). Positioning also remains challenging, with a
mean F1 of 32.22 and a best score of 45.35 (Qwen3.8-Max). Taken together, these results
indicate persistent gaps in component selection, property configuration, and layout
relationships in generated code, highlighting barriers to generating production-ready UI code.

\end{samepage}

\paragraph{Finding 2.} \emph{MLLM performance varies across requirement categories,
with alignment and positioning remaining key weaknesses.}
In T2, Alignment and Position average 29.94 and 45.98 F1, below Component (58.01) and
Group (56.91) (Table~\ref{tab:main}). Alignment also has the lowest overall mean in T3 with supplied
requirements (41.13 F1; Table~\ref{tab:guided}). These results suggest that MLLMs struggle
to reason about how elements relate to their containers and positioning references,
and to translate spatial constraints into layout code.

\begin{table*}[t]
\setlength{\belowcaptionskip}{9pt}
\caption{Results for Requirement Realization (T3) on 421 designs.
Values below each score show $\Delta=\mathrm{T3}-\mathrm{T1}$. Purple and yellow denote
gains and losses, respectively; bold marks the largest gain per column.}
\label{tab:guided}
\centering

\small
\setlength{\tabcolsep}{2.0pt}
\renewcommand{\arraystretch}{1.15}

\begin{tabular*}{\textwidth}{
@{\extracolsep{\fill}}
l
ccccc
ccc
@{}
}
\toprule
\multirow{2}{*}{Model}
& \multicolumn{5}{c}{\textbf{Code Implementation}}
& \multicolumn{3}{c}{\textbf{Visual Fidelity}}
\\
\cmidrule(lr){2-6}
\cmidrule(lr){7-9}

& Type
& Props
& Group
& Align.
& Pos.
& CV $\uparrow$
& VES $\uparrow$
& MAE $\downarrow$
\\
\midrule

GPT-5.6 Terra
& \rhi{17} \stk{81.72}{+19.25}
& \rhi{20} \stk{66.19}{+21.94}
& \rhi{4} \stk{61.11}{+4.27}
& \rhi{15} \stk{47.22}{+16.69}
& \rhi{13} \stk{51.03}{+14.60}
& \rhi{1} \stk{86.52}{+0.74}
& \rhi{1} \stk{\textbf{0.9263}}{+0.0089}
& \rhi{1} \stk{\textbf{0.1016}}{$-$0.0147}
\\

GPT-5.4
& \rhi{14} \stk{75.75}{+15.56}
& \rhi{19} \stk{61.44}{+21.15}
& \rhi{7} \stk{64.54}{+8.06}
& \rhi{20} \stk{49.44}{+21.96}
& \rhi{22} \stk{58.43}{+24.99}
& \stk{84.95}{$-$0.14}
& \rlo{1} \stk{0.9118}{$-$0.0065}
& \stk{0.1088}{+0.0010}
\\

Claude Sonnet 5
& \rhi{16} \stk{78.08}{+18.19}
& \rhi{21} \stk{63.99}{+23.62}
& \rhi{6} \stk{61.05}{+6.23}
& \rhi{10} \stk{35.16}{+11.62}
& \rhi{19} \stk{49.35}{+21.42}
& \rhi{1} \stk{79.40}{+0.58}
& \rhi{1} \stk{0.9003}{+0.0057}
& \stk{0.1344}{$-$0.0006}
\\

Qwen3.8-Max
& \rhi{16} \stk{79.99}{+18.22}
& \rhi{22} \stk{63.08}{+24.25}
& \rhi{5} \stk{64.79}{+5.33}
& \rhi{17} \stk{45.79}{+19.37}
& \rhi{18} \stk{64.82}{+19.47}
& \rlo{1} \stk{84.73}{$-$0.80}
& \rlo{1} \stk{0.9126}{$-$0.0066}
& \stk{0.1139}{+0.0005}
\\

Qwen3-VL-235B
& \rhi{10} \stk{42.83}{+11.23}
& \rhi{11} \stk{29.48}{+12.14}
& \stk{30.73}{+0.20}
& \rhi{3} \stk{9.81}{+3.09}
& \rhi{5} \stk{13.45}{+5.84}
& \stk{58.69}{+0.32}
& \rlo{1} \stk{0.7784}{$-$0.0135}
& \stk{0.1782}{$-$0.0026}
\\

Gemini 3.5 Flash
& \rhi{18} \stk{82.08}{+20.01}
& \rhi{27} \stk{67.49}{+29.46}
& \rhi{7} \stk{59.03}{+7.75}
& \rhi{15} \stk{37.00}{+16.90}
& \rhi{22} \stk{54.32}{+24.83}
& \stk{80.58}{+0.08}
& \rlo{2} \stk{0.8944}{$-$0.0168}
& \stk{0.1145}{$-$0.0050}
\\

Kimi K3
& \rhi{15} \stk{79.42}{+16.17}
& \rhi{21} \stk{64.64}{+23.08}
& \rhi{6} \stk{66.29}{+6.88}
& \rhi{18} \stk{46.81}{+19.77}
& \rhi{17} \stk{62.13}{+18.45}
& \rlo{1} \stk{83.54}{$-$0.86}
& \stk{0.9131}{$-$0.0020}
& \stk{0.1149}{$-$0.0029}
\\

Gemini 3.1 Pro
& \rhi{25} \stk{\textbf{85.80}}{+28.06}
& \rhi{34} \stk{\textbf{75.11}}{+38.24}
& \rhi{14} \stk{\textbf{71.34}}{+15.82}
& \rhi{30} \stk{\textbf{57.82}}{+33.23}
& \rhi{39} \stk{\textbf{77.11}}{+43.30}
& \rhi{1} \stk{\textbf{84.53}}{+1.53}
& \stk{0.9154}{$-$0.0001}
& \stk{0.1147}{$-$0.0019}
\\

\midrule
Mean
& \stk{75.71}{+18.34}
& \stk{61.43}{+24.24}
& \stk{59.86}{+6.82}
& \stk{41.13}{+17.83}
& \stk{53.83}{+21.61}
& \stk{80.37}{+0.18}
& \stk{0.8940}{$-$0.0039}
& \stk{0.1226}{$-$0.0033}
\\

\bottomrule
\end{tabular*}
\end{table*}

\paragraph{Finding 3.} \emph{MLLMs exhibit distinct performance profiles in requirement
inference and realization.}
GPT-5.6 Terra leads alignment inference at 37.11 F1, while Gemini 3.1 Pro ranks seventh
at 20.64 (Table~\ref{tab:main}). When expert requirements are supplied in T3, Gemini 3.1 Pro
leads alignment realization at 57.82, exceeding GPT-5.6 Terra's 47.22
(Table~\ref{tab:guided}). Grouping shows a similar contrast: Qwen3.8-Max leads inference
at 71.40 F1, whereas Gemini 3.1 Pro leads realization at 71.34.
These contrasting profiles suggest that MLLM improvement should target MLLM-specific
weaknesses: strengthening cross-modal reasoning over design images and layer metadata
for requirement inference, and the ability to translate component and layout
constraints into code for requirement realization.

\begin{figure}[t]
\centering
\includegraphics[width=\linewidth]{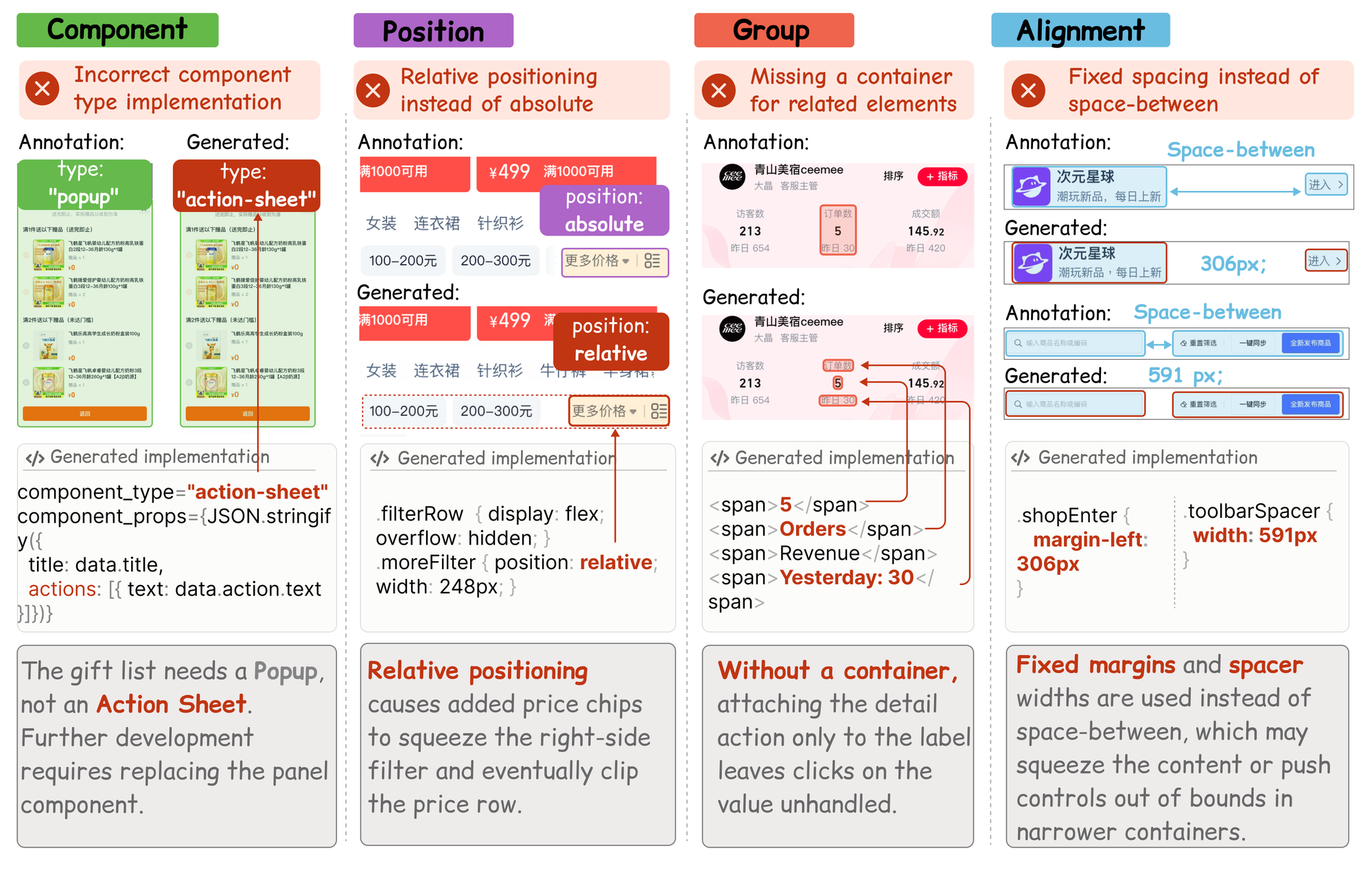}
\caption{\textbf{Visually similar interfaces can violate implementation requirements.}
From left to right: incorrect component type, incorrect positioning, missing grouping
container, and fixed spacing instead of space-between alignment. Code excerpts reveal
these errors; bottom panels explain their potential impact on component integration,
interaction handling, and layout adaptation.}
\label{fig:case}
\end{figure}

\paragraph{Finding 4.} \emph{MLLMs' visual reconstruction ability does not necessarily imply
an ability to generate UI code that satisfies implementation requirements.}
In T1, Kimi K3 ranks among the top three on all five code dimensions, but fourth on CV
and fifth on VES/MAE (Table~\ref{tab:main}). From T1 to T3, all MLLMs improve on every code
dimension (mean gains: 6.82--24.24 F1 points), while mean visual scores change little
(CV: +0.18; VES: $-$0.0039; MAE: $-$0.0033; Table~\ref{tab:guided}).
Figure~\ref{fig:case} further shows that generated code can contain errors even when the
rendered interfaces closely resemble the designs.
These results suggest limitations in relational abstraction and compositional reasoning
when translating visual designs into code.

\paragraph{Finding 5.} \emph{Requirements depend on the specific design context, and a
single design can allow multiple valid implementations.}
Absolute positioning is required for the filter in Figure~\ref{fig:case} but violates
normal-flow requirements for cards in another design, illustrating why heuristics favoring
less absolute positioning cannot determine its appropriateness.
The evaluation subset also contains 350 optional positioning, 170 optional alignment, and
49 optional grouping annotations, recording permitted implementation choices. For example,
the same design may allow either \texttt{justify-content: space-between} or
\texttt{margin-left: auto} to achieve the required spacing. A single-reference comparison
may penalize this difference even when both implementations satisfy the implementation requirements.

\begin{samepage}
\section{Conclusion}
\label{sec:conclusion}

TaoD2C-Bench pairs industrial UI designs with layer-level expert requirements to evaluate generated
code. Its three tasks reveal limited capabilities in both inferring implementation requirements and
implementing them in code, alongside distinct strengths across MLLMs and tasks.
These findings provide insights for future research on UI code generation.

\end{samepage}

\bibliography{iclr2027_conference}
\bibliographystyle{iclr2027_conference}

\appendix

\clearpage
\section*{Appendix Contents}
\noindent\begin{tabular*}{\linewidth}{@{}l@{\extracolsep{\fill}}r@{}}
\textbf{Section} & \textbf{Page} \\
\addlinespace[4pt]
\hyperref[app:platform]{\ref*{app:platform}\quad Dataset and Annotation} & \pageref{app:platform} \\
\quad Industrial diversity and complexity & \\
\quad Annotation platform, preprocessing, and quality control & \\
\quad Evaluation subset and scoring references & \\
\addlinespace[3pt]
\hyperref[app:prompts]{\ref*{app:prompts}\quad Task Prompts} & \pageref{app:prompts} \\
\quad Shared inputs, task instructions, and output formats & \\
\addlinespace[3pt]
\hyperref[app:metrics]{\ref*{app:metrics}\quad Evaluation Details} & \pageref{app:metrics} \\
\quad Requirement prediction and implementation, visual fidelity & \\
\quad Aggregation and failure handling & \\
\addlinespace[3pt]
\hyperref[app:inference]{\ref*{app:inference}\quad Experimental Settings} & \pageref{app:inference} \\
\quad Models, generation settings, and output validation & \\
\addlinespace[3pt]
\hyperref[app:domains]{\ref*{app:domains}\quad Domain-wise Results} & \pageref{app:domains} \\
\addlinespace[3pt]
\hyperref[app:license]{\ref*{app:license}\quad License} & \pageref{app:license} \\
\end{tabular*}
\medskip

\clearpage
\clearpage
\section{Dataset and Annotation}
\label{app:platform}

\subsection{Industrial Realism and Application Diversity}
\label{app:dataset-diversity}

TaoD2C brings together production designs from 17 commercial platforms, pairing the diversity
of real industrial interfaces with expert knowledge of their implementation requirements.
The following statistics cover all 2,861 final designs, rather than only the evaluation subset.
Table~\ref{tab:dataset-inventory} summarizes the scale of this resource.

\begin{table}[H]
\centering
\caption{Overall scale of TaoD2C.}
\label{tab:dataset-inventory}
\begin{tabular}{lr}
\toprule
Statistic & Count \\
\midrule
Commercial platforms & 17 \\
Designs & 2,861 \\
Visible layers & 183,697 \\
Expert annotation records & 97,652 \\
\bottomrule
\end{tabular}
\end{table}

\begin{figure}[H]
\centering
\includegraphics[width=\linewidth]{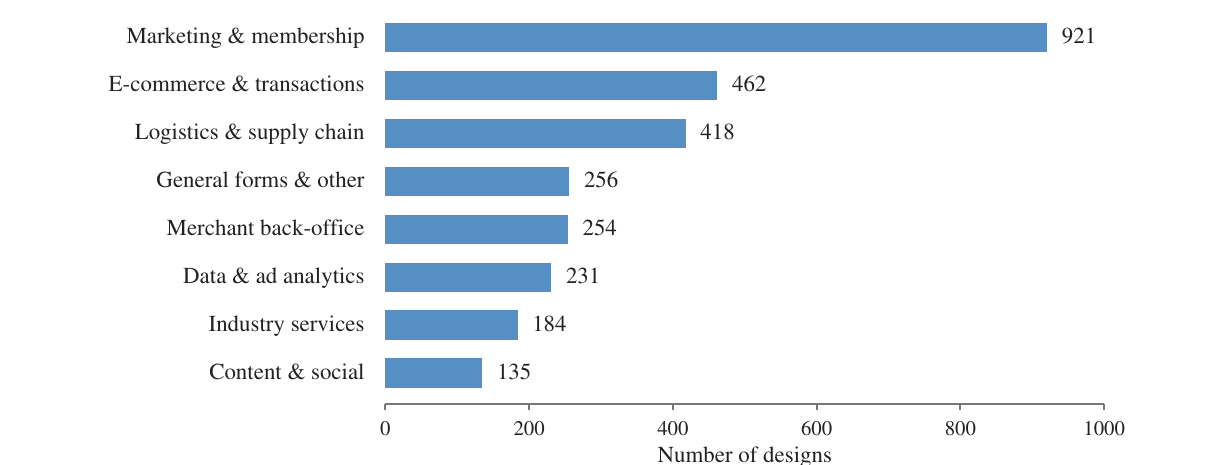}
\caption{Application scenarios across all 2,861 designs. Each design is counted once in its primary scenario.}
\label{fig:industrial-scenes}
\end{figure}

Figure~\ref{fig:industrial-scenes} summarizes eight application scenarios, including
marketing, transactions, logistics, merchant operations, and analytics. These scenarios
provide varied contexts for evaluating implementation requirement inference and realization.

\subsection{Structural Complexity and Annotation Coverage}
\label{app:structural-complexity}

Table~\ref{tab:dataset-complexity} summarizes visible layers and annotation records per design.
The median design contains 43 visible layers and 23 annotations; the corresponding 90th
percentiles are 146 and 81. These statistics describe the full dataset, rather than the
compiled scoring references for the evaluation subset.

\begin{table}[H]
\centering
\caption{Per-design statistics across all 2,861 designs. P90 denotes the 90th percentile.}
\label{tab:dataset-complexity}
\begin{tabular}{lrrrrr}
\toprule
Quantity & Mean & Median & P90 & Minimum & Maximum \\
\midrule
Visible layers & 64.21 & 43 & 146 & 1 & 776 \\
Expert annotation records & 34.13 & 23 & 81 & 0 & 270 \\
\bottomrule
\end{tabular}
\end{table}

\clearpage
\subsection{Annotation Platform}

\begin{figure}[H]
\centering
\includegraphics[width=\linewidth]{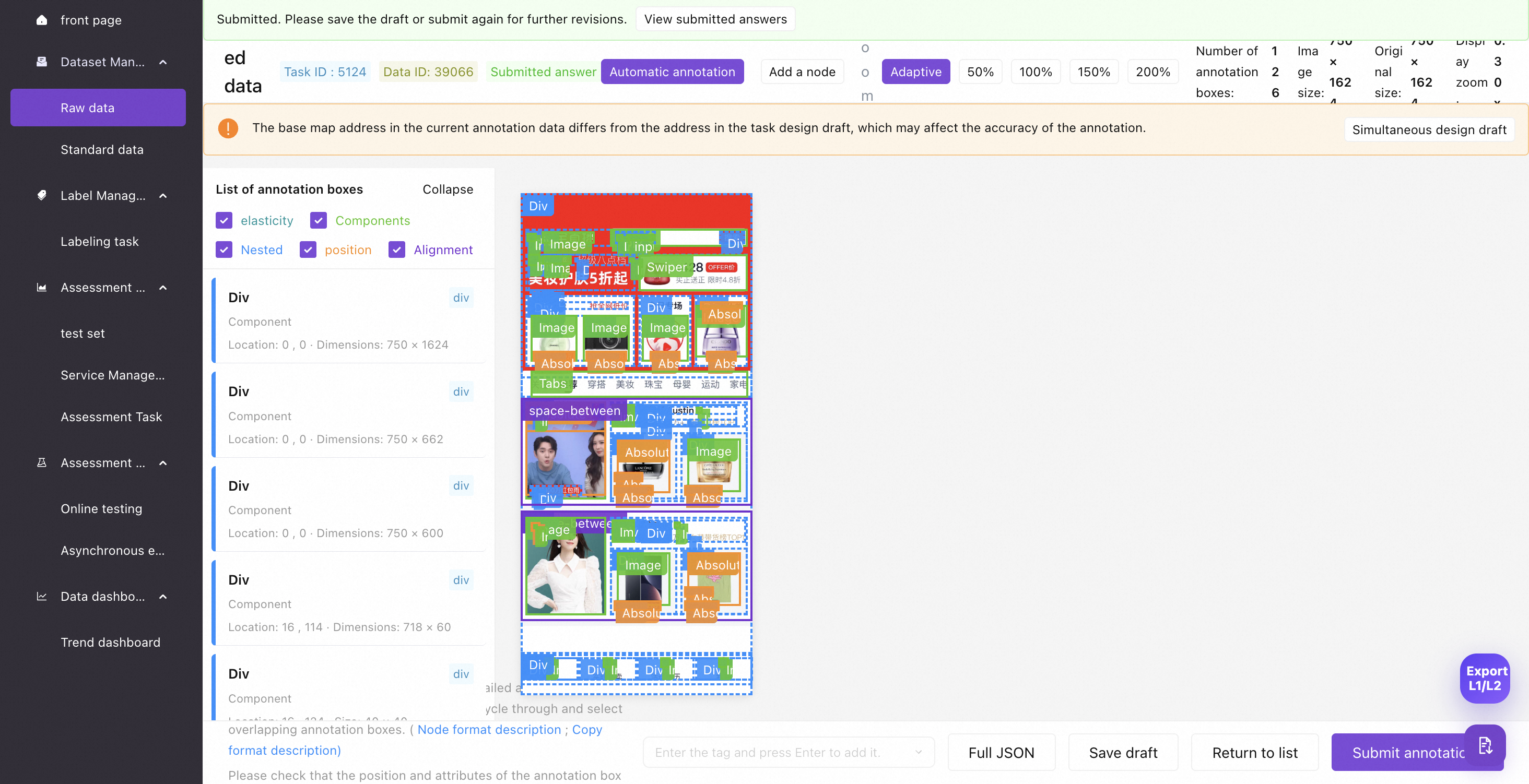}
\caption{The annotation platform. Experts inspect the design and its layers, overlay the four
requirement categories, and review individual annotations in the adjacent panel.}
\label{fig:platform}
\end{figure}

Experts annotate the original design alongside its rendered image and parsed layer tree
(Figure~\ref{fig:platform}). The platform links annotations to design layers and supports
Component, Group, Alignment, and Position requirements. Each entry is marked as required
or optional; acceptable alternative component types can also be recorded.

\subsection{Layer Parsing and Normalization}

Each design is parsed into a layer tree containing node identifiers, geometry, names,
component references, text, typography, visual styles, and image assets. Non-public links
and redundant artboard images are removed. Nodes with non-positive area remain in the
metadata but are excluded from member derivation and scoring. Each design is associated
with the target-library documentation used for annotation and evaluation. The annotated pool
is deduplicated by design URL, retaining the latest version of each design. This reduces
2,882 annotated design records to 2,861 final designs.

\subsection{Quality Control}

Annotations were produced by two senior frontend engineers with seven and five years of
experience, respectively. A third expert reviewed every annotated sample over multiple rounds
against the original design file, rendered image, and layer structure.
The review covered component types and properties, grouping, arrangement, and positioning,
and also checked for missing elements, position drift, and abnormal asset parsing. Samples with
errors were returned to the frontend experts for revision and rechecked before inclusion in the
final curated set. The annotation and verification process spanned 18 months.

\subsection{The Evaluation Subset}
\label{app:subset}

The evaluation subset comprises 421 designs from five annotation collections designated for
evaluation, including 152 B-end and 269 C-end designs. After removing six duplicate designs
and seven designs withdrawn during data preparation, the subset was frozen before MLLM
inference. Selection was performed at the collection level, without filtering individual
designs by content, complexity, or MLLM performance.

\subsection{Scoring Reference Construction}
\label{app:ref}

We convert expert annotations into a structured scoring reference $R^*$, linking each
annotation to its design region and corresponding layers. Each annotation has a field
indicating whether it is required or optional. Some optional annotations additionally
specify alternative implementations in an associated field. The resulting reference covers
Component, Group, Alignment, and Position and is shared across the three tasks: T1 and T3
evaluate generated code against it, while T2 evaluates predicted requirements. T3
additionally receives the expert annotations for each input design.
Table~\ref{tab:optional-counts} summarizes the required and optional reference entries in
the evaluation subset.

\begin{table}[htbp]
\centering
\caption{Required and optional reference entries in the 421-design evaluation subset.}
\label{tab:optional-counts}
\begin{tabular}{lrrr}
\toprule
Category & Required & Optional & Total \\
\midrule
Component & 7,976 & 21 & 7,997 \\
Group & 12,658 & 49 & 12,707 \\
Alignment & 747 & 170 & 917 \\
Position & 711 & 350 & 1,061 \\
\bottomrule
\end{tabular}
\end{table}

\clearpage
\section{Task Prompts}
\label{app:prompts}

Figures~\ref{fig:prompt-t1}--\ref{fig:prompt-t3} present prompt summaries for the three tasks.
Figure~\ref{fig:prompt-output} presents their output formats.

\subsection{Shared Inputs and Prompt Design}
\label{app:cache}

All three tasks receive design images, layer metadata, and target-library documentation,
which provide visual appearance, layer structure and attributes, and available components
and properties, respectively. T3 additionally receives the expert annotations for each input design.
Within each task, all MLLMs use the same prompt template.

\subsection{T1: End-to-End UI Code Generation}
\label{app:prompt-t1}

\begin{figure}[H]
\centering
\includegraphics[width=0.92\linewidth]{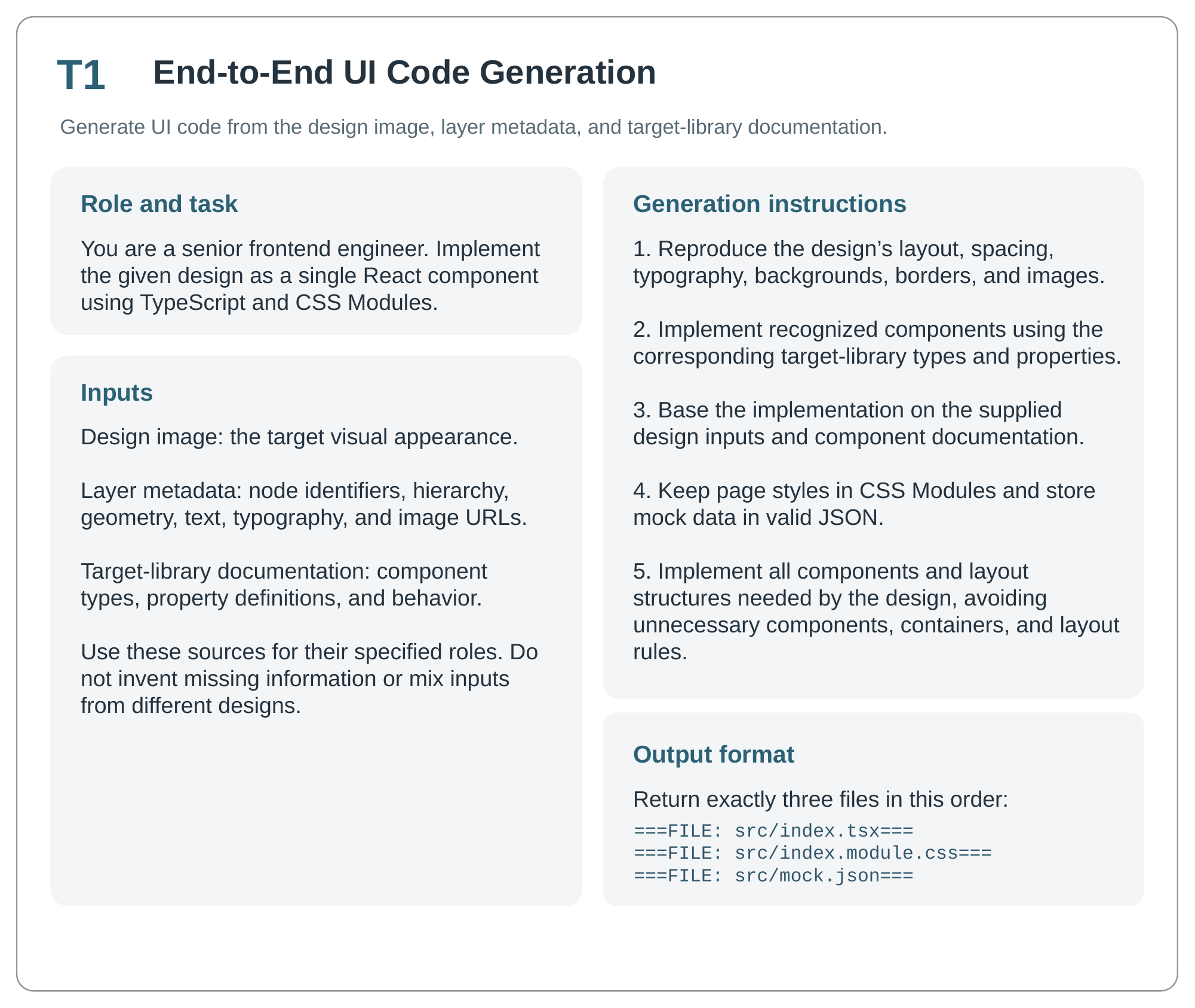}
\caption{End-to-End UI Code Generation (T1) prompt summary, including the shared
inputs, code output format, and generation constraints.}
\label{fig:prompt-t1}
\end{figure}

T1 generates a React page using TypeScript and CSS Modules from the shared design inputs.
Figure~\ref{fig:prompt-t1} summarizes the generation instructions, and
Figure~\ref{fig:prompt-output} specifies the three-file output format.

\clearpage
\subsection{T2: Requirement Inference}
\label{app:prompt-t2}

\begin{figure}[H]
\centering
\includegraphics[width=0.92\linewidth]{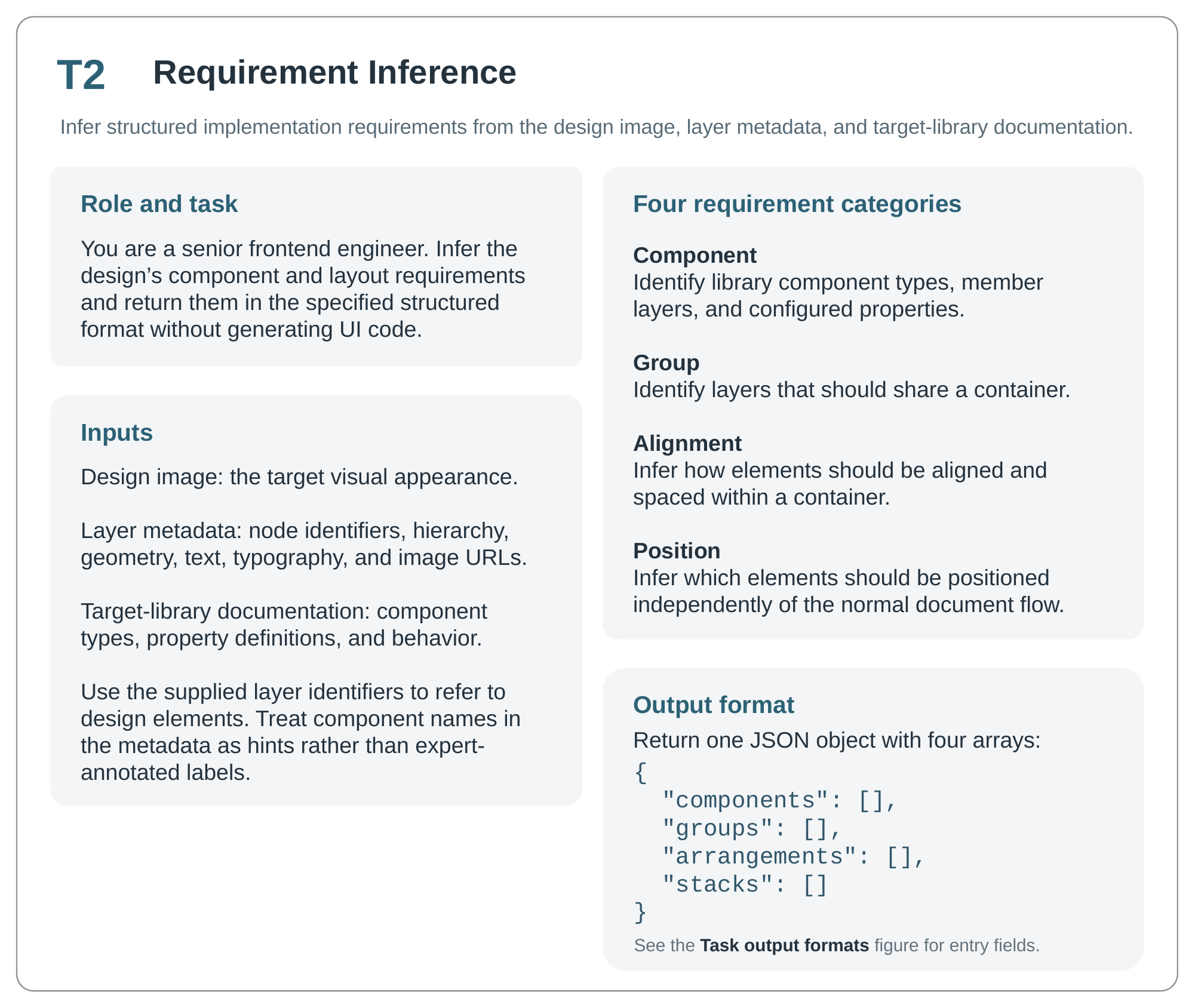}
\caption{Requirement Inference (T2) prompt summary.
The four requirement categories correspond to the components, groups, arrangements, and stacks arrays.}
\label{fig:prompt-t2}
\end{figure}

T2 infers structured implementation requirements from the shared design inputs without generating
code. Figure~\ref{fig:prompt-t2} summarizes the task instructions, and
Figure~\ref{fig:prompt-output} specifies the output schema.

\clearpage
\subsection{T3: Requirement Realization}
\label{app:prompt-t3}

\begin{figure}[H]
\centering
\includegraphics[width=0.92\linewidth]{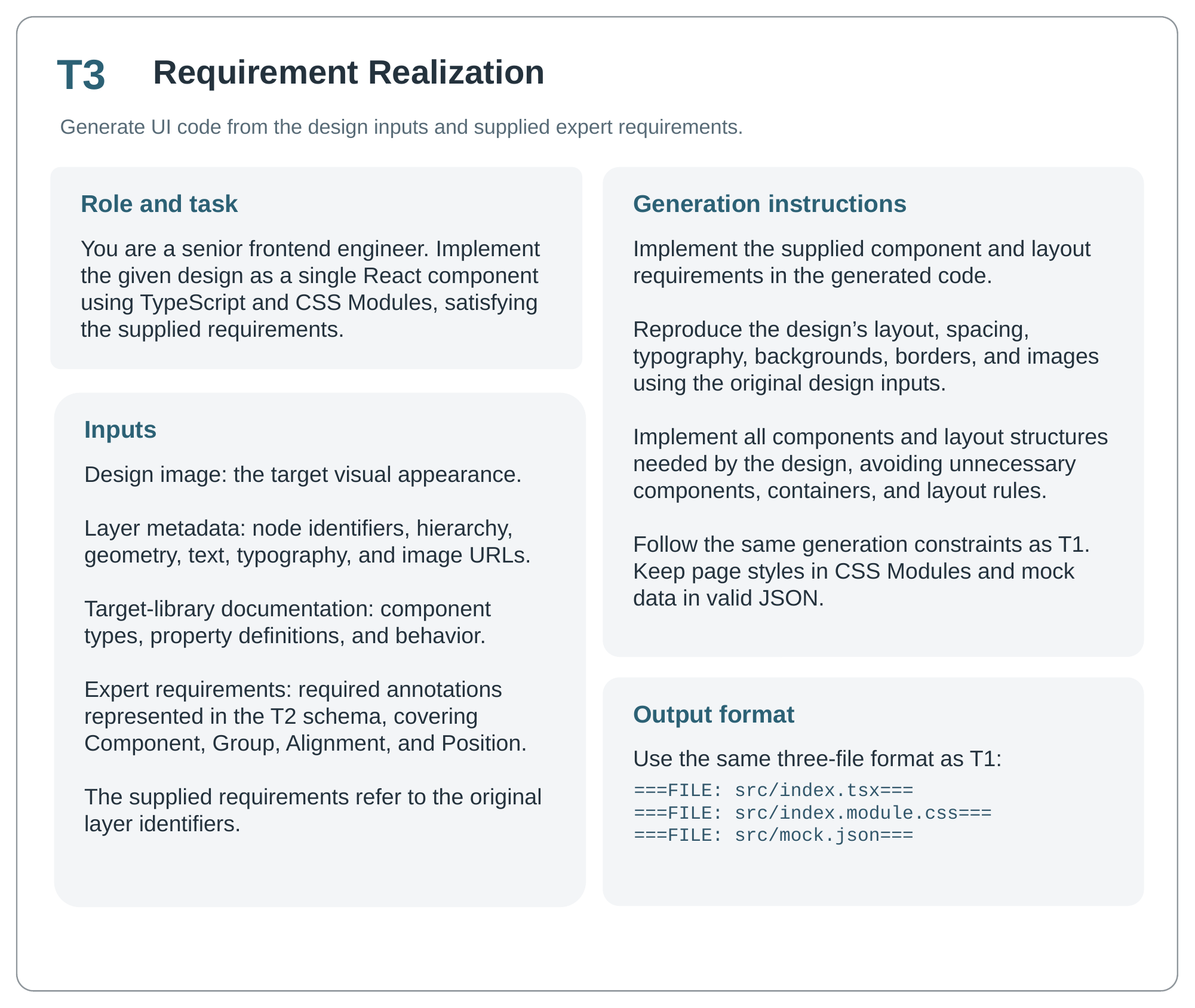}
\caption{Requirement Realization (T3) prompt summary.
T3 supplements the T1 inputs with the expert annotations for the input design while retaining
the same code generation instructions and output format.}
\label{fig:prompt-t3}
\end{figure}

T3 extends T1 with the expert annotations for each input design
(Figure~\ref{fig:prompt-t3}). The MLLM implements the supplied requirements while using
the original design inputs for text, images, geometry, and visual styles. Code output,
rendering, and evaluation remain the same as in T1.

\clearpage
\subsection{Output Formats}
\label{app:output-formats}

\begin{figure}[H]
\centering
\includegraphics[width=0.92\linewidth]{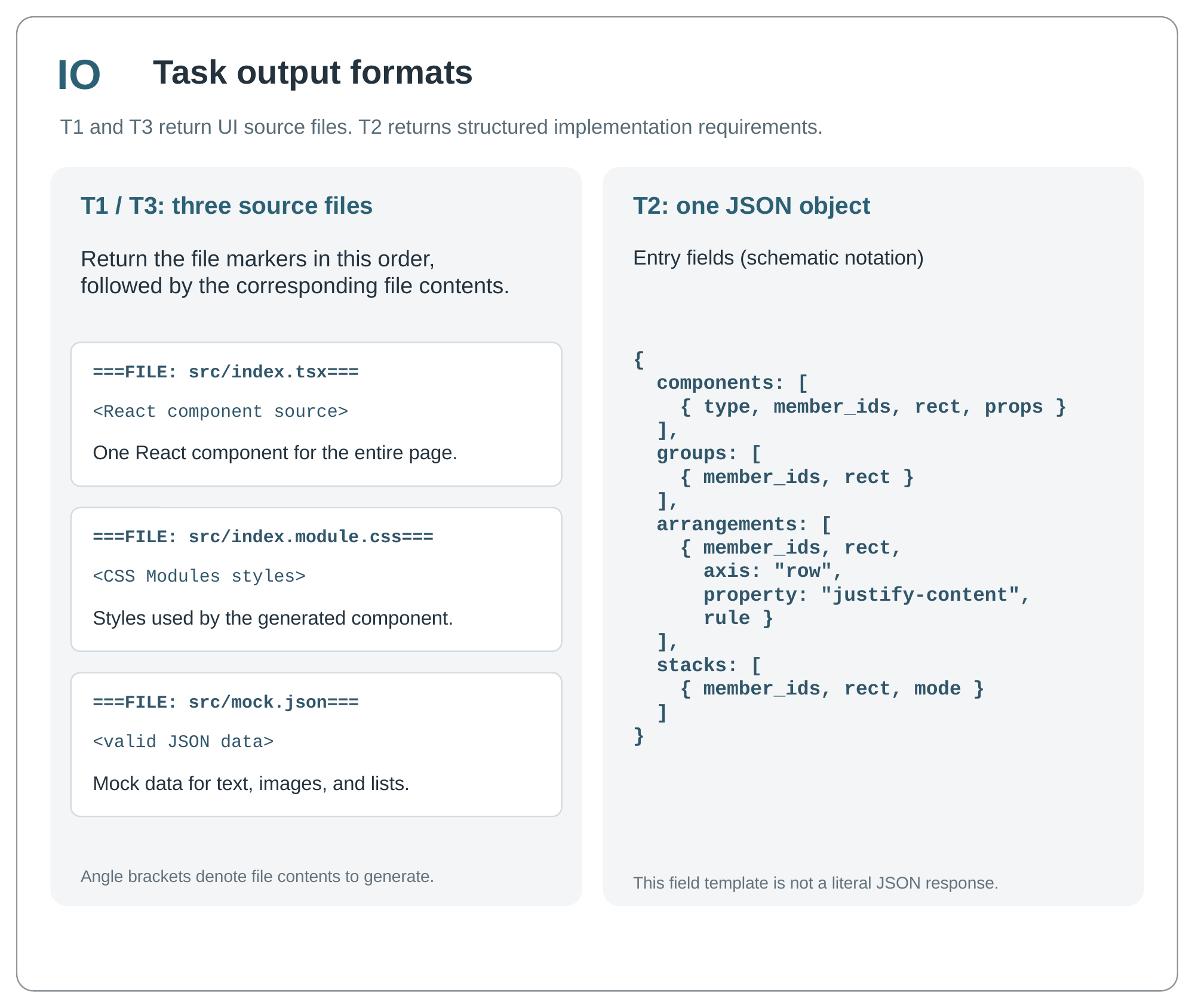}
\caption{Output formats for the three TaoD2C-Bench tasks. Left: ordered file blocks for
T1 and T3. Right: the four arrays and their entry fields for T2. File contents and field notation are schematic,
not an executable response or a complete JSON example.}
\label{fig:prompt-output}
\end{figure}

Figure~\ref{fig:prompt-output} summarizes the output contracts. T1 and T3 return
three source files delimited by fixed file markers, whereas T2 returns one JSON
object containing four arrays. The figure uses placeholders for generated file
contents and schematic notation for JSON entry fields. Actual T2 responses contain
valid JSON values and follow the output schema in Figure~\ref{fig:prompt-output}. The expert requirements supplied to T3 follow the same
public schema as the T2 output.

\clearpage
\section{Evaluation Details}
\label{app:metrics}

Requirement Prediction and Code Implementation use the same four requirement categories
but evaluate different outputs. The former evaluates T2's structured requirements; the latter
evaluates their implementation in T1/T3 code and its runtime DOM and CSS. Code Implementation
reports component Type and Props separately, giving five F1 scores rather than four.
Both use micro-averaged precision, recall, and F1, with F1 reported in the tables.

For a dimension, let $\mathcal{P}_s$ be the candidate predictions of sample $s$,
$\mathcal{R}^{\mathrm{req}}_s$ its required reference entries, $M_s$ the number of matched
candidates, and $M^{\mathrm{req}}_s$ the number of matched required entries. Counts are pooled
across all samples before division:
\begin{equation}
\label{eq:microprf}
\mathrm{P}=\frac{\sum_s M_s}{\sum_s |\mathcal{P}_s|},\qquad
\mathrm{R}=\frac{\sum_s M^{\mathrm{req}}_s}{\sum_s |\mathcal{R}^{\mathrm{req}}_s|},\qquad
\mathrm{F1}=\frac{2\,\mathrm{P}\,\mathrm{R}}{\mathrm{P}+\mathrm{R}}.
\end{equation}

\subsection{Requirement Prediction Scorer}
\label{app:dl}

The scorer deterministically matches T2 predictions to expert annotations in each category.
A match requires identical, non-empty derived member sets and agreement on the relevant
attributes: component type for Component; axis, property, and rule for Alignment; and
positioning mode for Position. Group requires member-set agreement only.

\paragraph{Matching and optional entries.}
One-to-one maximum matching is performed first against required entries and then against
optional entries using the remaining predictions. Thus, $M=M^{\mathrm{req}}+M^{\mathrm{opt}}$
in Eq.~\ref{eq:microprf}. Unmatched predictions count as false positives; unmatched required
entries count as false negatives. Optional matches contribute to precision but not recall,
and omitting an optional entry incurs no penalty.

\paragraph{Member sets and evaluation scope.}
Predicted component member sets are expanded to include descendants in the layer tree.
For Group, Alignment, and Position, members are derived from predicted rectangles using
the same procedure as the reference (Appendix~\ref{app:ref}). The component evaluation
excludes \texttt{image}, \texttt{icon}, and \texttt{avatar} from both predictions and references.

\subsection{Code Implementation Scorer}
\label{app:codef1}

The scorer matches expert annotations to evidence from generated code and the rendered
DOM. Generated pages are bundled with esbuild and rendered in Playwright. Runtime evidence
includes bounding boxes, parent--child relations, computed styles, and component types and
properties associated with visual component roots. For two axis-aligned
boxes $a$ and $b$ with areas $|a|$ and $|b|$, the geometric tests use intersection-over-union
and area ratio,
\begin{equation}
\label{eq:iou}
\mathrm{IoU}(a,b)=\frac{|a\cap b|}{|a\cup b|},\qquad
\mathrm{AR}(a,b)=\frac{\min(|a|,|b|)}{\max(|a|,|b|)}.
\end{equation}

\paragraph{Type and Props.} Component types and properties are associated with their visual roots to define the
runtime boundaries used for matching. A runtime component implementation $u$ and a reference component $c$ form a legal edge when their
component types agree under the target-library alias relation $\equiv$ and both their IoU
and area similarity are at least $0.7$,
\begin{equation}
\label{eq:compedge}
(u,c)\in E_{\mathrm{comp}}\iff \mathrm{type}(u)\equiv\mathrm{type}(c)\ \wedge\ \mathrm{IoU}(u,c)\ge 0.7\ \wedge\ \mathrm{AR}(u,c)\ge 0.7.
\end{equation}
Matching is one-to-one and maximizes the number of legal pairs. Component names are
normalized using target-library-specific aliases. Plain containers and image types are
excluded from component evaluation.
Props are checked only on the component implementations whose type is already correct, and
the two are reported as separate Type and Props columns.

\paragraph{Group, Alignment, and Position.} Group candidates are containers with at least two visible children, or single-child
containers with a \texttt{data-grounding-group} marker or geometric agreement with a
single-member reference. In the latter case, the matched member node must be a descendant
of the container when available. All candidates enter the precision denominator and must
satisfy the same geometric matching criteria. Matching between a reference group $g$
and a candidate container $c$ requires both intersection-over-union and area similarity
to be at least $0.7$. Area similarity is measured by
$\mathrm{AR}(g,c)$ as defined above.
A group with an empty member set has no structural identity
and is excluded from the shared required denominator, with excluded counts recorded.
Alignment requires both container $\mathrm{IoU}\ge 0.7$ and the same alignment type
as the annotation. Position pairs at $\mathrm{IoU}\ge 0.7$ and checks the positioning mode.
These dimensions use one-to-one matching that maximizes the number of matches,
with geometric scores used to resolve ties. Runtime bounding boxes are calibrated by the
median anchor offset before the geometric tests, removing a global page shift without
changing relative geometry.

\subsection{Visual Fidelity}
\label{app:visual}

Let $I$ be the design image and $I_C=\mathrm{Render}(C)$ the rendered image. We report CV, VES, and MAE
separately from Code Implementation and do not combine the three into an overall
score. 
\paragraph{Composite visual score (CV).}
We adopt the visual similarity metric in Section~3.3 and Appendix~B of
1D-Bench \citep{xu2026onedbench}. It combines perceptual image similarity
$S_{\mathrm{img}}$, element completeness $S_{\mathrm{comp}}$, and layout similarity
$S_{\mathrm{layout}}$, each on $[0,1]$. We report CV on a $0$--$100$ scale:
\begin{equation}
\label{eq:cv}
\mathrm{CV}(I,I_C)=100\big(0.5\,S_{\mathrm{img}}+0.3\,S_{\mathrm{comp}}
+0.2\,S_{\mathrm{layout}}\big).
\end{equation}
The CV preprocessing handles transparency and aligns both images to width
$w=\min(w_I,w_{I_C})$ and height $h=\min(h_I,h_{I_C})$ before detection and comparison.

We follow its text and non-text matching procedures. Unavailable signals are omitted
and the remaining weights are renormalized; when neither text nor non-text blocks
are detected, CV reduces to $100\,S_{\mathrm{img}}$. The pixel-difference term within
CV is distinct from the separately reported MAE below.

\paragraph{Visual element similarity (VES).}
VES compares the CLS-token embeddings produced by a DINOv2-base encoder $\phi$ \citep{oquab2023dinov2}:
\begin{equation}
\label{eq:ves}
\mathrm{VES}(I,I_C)=\frac{\phi(I)^{\mathsf T}\phi(I_C)}
{\|\phi(I)\|_2\,\|\phi(I_C)\|_2}.
\end{equation}
Higher CV and VES indicate better visual fidelity.

\paragraph{Mean absolute error (MAE).}
For MAE, we crop the near-white borders of both images at threshold $250$, resize
the render proportionally to the reference, and center-pad it with white.
Let $\tilde I,\widetilde{I_C}\in[0,255]^{H\times W\times 3}$ denote the resulting RGB images.
The normalized error is
\begin{equation}
\label{eq:mae}
\mathrm{MAE}(I,I_C)=\frac{1}{255\cdot 3HW}
\sum_{i=1}^{H}\sum_{j=1}^{W}\sum_{k=1}^{3}
\left|\tilde I_{ijk}-(\widetilde{I_C})_{ijk}\right|.
\end{equation}
Lower MAE indicates better visual fidelity. All three metrics are reported as case
means using one final output per design; failure handling and aggregation follow
Appendix~\ref{app:protocol}.

\subsection{Aggregation and Failure Handling}
\label{app:protocol}
All eight models are evaluated on the same 421 designs for all three tasks, yielding
10,104 task outcomes. Results use the final output after the retries and repairs described
in Appendix~\ref{app:prompt-execution}, rather than first-attempt performance.
A successful repair replaces the failed attempt.

There are 10,101 successful outcomes and three failures: all three tasks for
Qwen3-VL-235B on B-end design 629 exceed the provider's 126,976-token input limit.
Failures receive F1, CV, and VES of zero and MAE of one, where applicable. Required
reference entries remain in the micro-recall denominator with zero matches; dimensions
without reference entries receive no artificial reference count.

Requirement scores pool counts across designs as defined in Eq.~\ref{eq:microprf}.
CV, VES, and MAE are averaged across designs. Table Mean rows average model scores;
they do not pool counts across models. Model means and T3--T1 differences use the
reported precision: two decimals for F1 and CV, and four for VES and MAE.

\section{Experimental Settings}
\label{app:inference}

\subsection{Models and Generation Settings}
\label{app:models}
Table~\ref{tab:model-settings} lists model-specific inference settings. GPT-5.6 Terra
and GPT-5.4 use Codex CLI; the other models use an OpenAI-compatible gateway.
The same 421 designs and task templates are used across all models.

\begin{table}[htbp]
\centering
\small
\caption{Model-specific inference settings. A dash indicates that no setting
is specified. HTTP caps are transport limits, not task-level budgets.}
\label{tab:model-settings}
\begin{tabular}{lll}
\toprule
Model & Reasoning setting & HTTP output cap \\
\midrule
GPT-5.6 Terra & Medium & --- \\
GPT-5.4 & Medium & --- \\
Claude Sonnet 5 & Adaptive, medium effort & 60,000 \\
Qwen3.8-Max & --- & 60,000 \\
Qwen3-VL-235B-A22B-Thinking & --- & 32,768 \\
Gemini 3.5 Flash & Medium & 60,000 \\
Kimi K3 & Low & 60,000 \\
Gemini 3.1 Pro Preview & --- & 60,000 \\
\bottomrule
\end{tabular}
\end{table}

The task runner initially requests 32,000 output tokens. A response truncated at the
length limit is retried once with a 48,000-token budget when the model's ceiling permits.
These budgets are subject to the transport limits in Table~\ref{tab:model-settings};
the Codex CLI transport does not pass an output-token cap.
Qwen3.8-Max uses two gateway entries, \texttt{Qwen3.8-Max} and \texttt{Qwen0902},
for the same model; results are reported under its canonical name.

\subsection{Execution and Output Validation}
\label{app:client}
\label{app:manifest}
\label{app:prompt-execution}

Requests contain the shared design inputs followed by the task instructions. T3 inserts
its expert requirements after the layer tree. Supported models use prompt caching;
cache controls do not change the task content. Experiment manifests record the input
samples, component documentation, prompts, and references used for each run.

\paragraph{Retries and repair.}
Transient request failures allow at most five retries after the initial attempt.
Deterministic request errors fail immediately. A timeout can trigger an identical
request using streaming. Separately, invalid output triggers one repair request
containing the original prompt and validation error. For T2, validation checks JSON
parsing and schema compliance; a surrounding code fence is accepted. For T1/T3, repair
is triggered by absent file blocks or a missing required file. Predictions are not
manually edited, and retry outputs are not selected by their evaluation scores.

\paragraph{Rendering and evaluation.}
Generated pages are built and rendered at the design's root width. Failed renders retain
an error record. T2 predictions are evaluated against the compiled reference;
T3 uses the same rendering and evaluation as T1. The annotations supplied to T3 are recorded for each input design.
Failure scoring and aggregation are specified in Appendix~\ref{app:protocol}.

\clearpage
\section{Domain-wise Results}
\label{app:domains}
Tables~\ref{tab:b-t1}--\ref{tab:c-t3} give all three tasks separately for the 152 B-end
and 269 C-end designs. Each domain uses the same cases across all eight models, with
terminal failures retained. F1 is micro-aggregated within each domain; the B+C F1 is
pooled from counts and is not an average of the B-end and C-end F1 values. CV, VES,
and MAE are case means. Model abbreviations follow Table~\ref{tab:main}.


\begin{table}[!htbp]
\centering
\small
\caption{B-end results for End-to-End UI Code Generation (T1) on the same 152 designs for all eight models. F1 is in percent. Bold marks the best value per column; Mean averages models. CV and VES are higher-better; MAE is lower-better.}
\label{tab:b-t1}
\setlength{\tabcolsep}{3pt}
\renewcommand{\arraystretch}{1.2}
\begin{tabular*}{\textwidth}{@{\extracolsep{\fill}}lcccccccc@{}}
\toprule
Model & Type & Props & Group & Align. & Pos. & CV $\uparrow$ & VES $\uparrow$ & MAE $\downarrow$ \\
\midrule
GPT-5.6 Terra & 67.74 & \textbf{47.21} & \textbf{41.87} & \textbf{39.95} & 23.63 & 82.23 & 0.9225 & 0.1221 \\
GPT-5.4 & 66.75 & 43.45 & 39.78 & 37.90 & 12.56 & \textbf{83.12} & \textbf{0.9408} & \textbf{0.0991} \\
Claude Sonnet 5 & 65.04 & 41.73 & 39.19 & 29.63 & 16.70 & 76.03 & 0.9022 & 0.1462 \\
Qwen3.8-Max & \textbf{69.11} & 40.57 & 39.03 & 31.47 & 16.97 & 79.68 & 0.9240 & 0.1352 \\
Qwen3-VL-235B & 30.67 & 13.95 & 26.53 & 9.65 & 11.38 & 61.40 & 0.8706 & 0.1824 \\
Gemini 3.5 Flash & 66.47 & 38.51 & 33.08 & 25.98 & 24.70 & 77.82 & 0.9278 & 0.1130 \\
Kimi K3 & 68.51 & 43.70 & 40.10 & 36.58 & \textbf{28.00} & 81.06 & 0.9258 & 0.1231 \\
Gemini 3.1 Pro & 68.04 & 41.56 & 37.77 & 32.21 & 25.39 & 80.14 & 0.9285 & 0.1177 \\
\midrule
Mean & 62.79 & 38.84 & 37.17 & 30.42 & 19.92 & 77.69 & 0.9178 & 0.1299 \\
\bottomrule
\end{tabular*}
\end{table}

\begin{table}[!htbp]
\centering
\small
\caption{B-end results for Requirement Inference (T2) on the same 152 designs for all eight models. F1 is in percent. Bold marks the best value per column; Mean averages models.}
\label{tab:b-t2}
\setlength{\tabcolsep}{3pt}
\renewcommand{\arraystretch}{1.2}
\begin{tabular*}{\textwidth}{@{\extracolsep{\fill}}lcccc@{}}
\toprule
Model & Comp. & Group & Align. & Pos. \\
\midrule
GPT-5.6 Terra & 59.70 & 56.94 & \textbf{38.17} & 25.19 \\
GPT-5.4 & 59.60 & 57.59 & 29.88 & \textbf{26.48} \\
Claude Sonnet 5 & 58.68 & 56.51 & 35.56 & 23.87 \\
Qwen3.8-Max & 59.97 & 61.16 & 24.15 & 21.57 \\
Qwen3-VL-235B & 53.54 & 37.00 & 21.91 & 16.05 \\
Gemini 3.5 Flash & 59.68 & 52.51 & 32.22 & 25.59 \\
Kimi K3 & \textbf{60.02} & \textbf{62.82} & 32.24 & 17.31 \\
Gemini 3.1 Pro & 59.47 & 60.37 & 17.73 & 16.39 \\
\midrule
Mean & 58.83 & 55.61 & 28.98 & 21.56 \\
\bottomrule
\end{tabular*}
\end{table}

\begin{table}[!htbp]
\centering
\small
\caption{B-end results for Requirement Realization (T3) on the same 152 designs for all eight models. F1 is in percent. Bold marks the best value per column; Mean averages models. CV and VES are higher-better; MAE is lower-better.}
\label{tab:b-t3}
\setlength{\tabcolsep}{3pt}
\renewcommand{\arraystretch}{1.2}
\begin{tabular*}{\textwidth}{@{\extracolsep{\fill}}lcccccccc@{}}
\toprule
Model & Type & Props & Group & Align. & Pos. & CV $\uparrow$ & VES $\uparrow$ & MAE $\downarrow$ \\
\midrule
GPT-5.6 Terra & 85.05 & 67.94 & \textbf{48.32} & 75.44 & 60.36 & \textbf{85.36} & \textbf{0.9447} & \textbf{0.0897} \\
GPT-5.4 & 80.78 & 66.72 & 43.69 & 66.27 & 62.90 & 84.26 & 0.9383 & 0.1013 \\
Claude Sonnet 5 & 83.04 & 66.09 & 42.30 & 55.25 & 70.23 & 78.47 & 0.9189 & 0.1400 \\
Qwen3.8-Max & 81.32 & 60.91 & 39.18 & 34.47 & 36.69 & 78.87 & 0.9226 & 0.1294 \\
Qwen3-VL-235B & 39.43 & 23.97 & 29.23 & 8.24 & 12.33 & 61.83 & 0.8545 & 0.1787 \\
Gemini 3.5 Flash & 85.81 & 70.53 & 39.54 & 56.72 & 65.62 & 78.34 & 0.9333 & 0.0981 \\
Kimi K3 & 83.08 & 67.95 & 46.13 & 67.36 & 68.91 & 81.28 & 0.9315 & 0.1182 \\
Gemini 3.1 Pro & \textbf{89.30} & \textbf{78.53} & 47.66 & \textbf{77.01} & \textbf{90.16} & 81.87 & 0.9315 & 0.1071 \\
\midrule
Mean & 78.48 & 62.83 & 42.01 & 55.10 & 58.40 & 78.79 & 0.9219 & 0.1203 \\
\bottomrule
\end{tabular*}
\end{table}

\clearpage

\begin{table}[!htbp]
\centering
\small
\caption{C-end results for End-to-End UI Code Generation (T1) on the same 269 designs for all eight models. F1 is in percent. Bold marks the best value per column; Mean averages models. CV and VES are higher-better; MAE is lower-better.}
\label{tab:c-t1}
\setlength{\tabcolsep}{3pt}
\renewcommand{\arraystretch}{1.2}
\begin{tabular*}{\textwidth}{@{\extracolsep{\fill}}lcccccccc@{}}
\toprule
Model & Type & Props & Group & Align. & Pos. & CV $\uparrow$ & VES $\uparrow$ & MAE $\downarrow$ \\
\midrule
GPT-5.6 Terra & 53.45 & \textbf{39.19} & 62.44 & \textbf{28.08} & 38.34 & 87.78 & 0.9145 & 0.1129 \\
GPT-5.4 & 48.93 & 34.88 & 63.52 & 25.24 & 38.41 & 86.20 & 0.9056 & 0.1127 \\
Claude Sonnet 5 & 51.67 & 38.20 & 60.99 & 22.10 & 29.55 & 80.40 & 0.8903 & 0.1287 \\
Qwen3.8-Max & 49.85 & 36.01 & \textbf{67.89} & 25.19 & \textbf{48.66} & \textbf{88.83} & \textbf{0.9165} & \textbf{0.1010} \\
Qwen3-VL-235B & 33.10 & 22.81 & 32.11 & 6.24 & 6.73 & 56.66 & 0.7475 & 0.1799 \\
Gemini 3.5 Flash & \textbf{54.51} & 37.21 & 58.91 & 18.71 & 30.03 & 82.02 & 0.9019 & 0.1232 \\
Kimi K3 & 54.25 & 37.90 & 67.28 & 24.63 & 45.04 & 86.29 & 0.9090 & 0.1148 \\
Gemini 3.1 Pro & 43.45 & 30.36 & 62.80 & 22.64 & 34.60 & 84.61 & 0.9081 & 0.1161 \\
\midrule
Mean & 48.65 & 34.57 & 59.49 & 21.60 & 33.92 & 81.60 & 0.8867 & 0.1237 \\
\bottomrule
\end{tabular*}
\end{table}

\begin{table}[!htbp]
\centering
\small
\caption{C-end results for Requirement Inference (T2) on the same 269 designs for all eight models. F1 is in percent. Bold marks the best value per column; Mean averages models.}
\label{tab:c-t2}
\setlength{\tabcolsep}{3pt}
\renewcommand{\arraystretch}{1.2}
\begin{tabular*}{\textwidth}{@{\extracolsep{\fill}}lcccc@{}}
\toprule
Model & Comp. & Group & Align. & Pos. \\
\midrule
GPT-5.6 Terra & 58.23 & 61.98 & \textbf{36.85} & 50.87 \\
GPT-5.4 & 53.47 & 61.87 & 32.61 & 52.32 \\
Claude Sonnet 5 & 57.81 & 48.90 & 34.81 & 50.59 \\
Qwen3.8-Max & 57.02 & \textbf{73.90} & 32.18 & 53.28 \\
Qwen3-VL-235B & 52.86 & 17.23 & 15.94 & 28.14 \\
Gemini 3.5 Flash & 57.27 & 51.18 & 35.39 & 50.11 \\
Kimi K3 & \textbf{59.09} & 72.83 & 33.82 & \textbf{56.01} \\
Gemini 3.1 Pro & 57.27 & 69.90 & 21.90 & 53.12 \\
\midrule
Mean & 56.63 & 57.22 & 30.44 & 49.31 \\
\bottomrule
\end{tabular*}
\end{table}

\begin{table}[!htbp]
\centering
\small
\caption{C-end results for Requirement Realization (T3) on the same 269 designs for all eight models. F1 is in percent. Bold marks the best value per column; Mean averages models. CV and VES are higher-better; MAE is lower-better.}
\label{tab:c-t3}
\setlength{\tabcolsep}{3pt}
\renewcommand{\arraystretch}{1.2}
\begin{tabular*}{\textwidth}{@{\extracolsep{\fill}}lcccccccc@{}}
\toprule
Model & Type & Props & Group & Align. & Pos. & CV $\uparrow$ & VES $\uparrow$ & MAE $\downarrow$ \\
\midrule
GPT-5.6 Terra & 75.56 & 62.95 & 65.81 & 41.47 & 50.17 & 87.17 & \textbf{0.9158} & 0.1083 \\
GPT-5.4 & 66.89 & 52.14 & 72.85 & 46.68 & 58.05 & 85.34 & 0.8968 & 0.1130 \\
Claude Sonnet 5 & 69.56 & 60.39 & 68.06 & 31.94 & 47.48 & 79.93 & 0.8898 & 0.1312 \\
Qwen3.8-Max & 77.56 & 67.01 & 74.90 & 50.00 & 68.02 & \textbf{88.04} & 0.9070 & \textbf{0.1050} \\
Qwen3-VL-235B & 49.53 & 40.37 & 31.32 & 10.08 & 13.57 & 56.92 & 0.7354 & 0.1779 \\
Gemini 3.5 Flash & 75.18 & 61.89 & 66.70 & 33.90 & 53.38 & 81.84 & 0.8724 & 0.1238 \\
Kimi K3 & 72.64 & 58.53 & 74.13 & 43.53 & 61.61 & 84.81 & 0.9027 & 0.1130 \\
Gemini 3.1 Pro & \textbf{79.66} & \textbf{69.08} & \textbf{80.37} & \textbf{54.88} & \textbf{75.95} & 86.03 & 0.9063 & 0.1189 \\
\midrule
Mean & 70.82 & 59.05 & 66.77 & 39.06 & 53.53 & 81.26 & 0.8783 & 0.1239 \\
\bottomrule
\end{tabular*}
\end{table}

\clearpage

\clearpage
\section{License}
\label{app:license}

All designs in TaoD2C originate from the production archives of 17 commercial
platforms, which retain copyright over the original design assets. The publicly available
evaluation subset is released for non-commercial research use under the two sets of terms below.

\paragraph{Benchmark-created materials.}
The expert annotations and compiled implementation requirements included in the public
evaluation subset are distributed under the Creative Commons
Attribution-NonCommercial 4.0 International (CC BY-NC 4.0) license. This license
permits sharing and adaptation for non-commercial purposes, provided that appropriate
credit is given and changes are indicated.

\paragraph{Platform-owned design assets.}
The original design files, rendered design images, extracted assets, and layer metadata
remain the property of the originating platforms. Materials included in the public evaluation
subset are provided under the accompanying non-commercial research data-use terms and may not be
redistributed or sublicensed separately. Commercial use of these platform-owned assets
requires prior permission from the originating platforms.

\end{document}